\documentclass[12pt,a4paper]{article}
\usepackage{jheppub}
\usepackage{epstopdf}
\usepackage{graphicx}
\usepackage{epsfig}
\usepackage{dcolumn}  
\usepackage{bm}    
\usepackage{amssymb} 
\usepackage{amsmath,bm}
\usepackage{amsfonts}    
\usepackage{slashed}  
\usepackage{youngtab}
\usepackage[mathscr]{euscript}
\usepackage{epsfig}
\usepackage{verbatim}
\usepackage[all]{xy}

\newdimen\tableauside\tableauside=1.0ex
\newdimen\tableaurule\tableaurule=0.4pt
\newdimen\tableaustep
\def\phantomhrule#1{\hbox{\vbox to0pt{\hrule height\tableaurule width#1\vss}}}
\def\phantomvrule#1{\vbox{\hbox to0pt{\vrule width\tableaurule height#1\hss}}}
\def\sqr{\vbox{%
		\phantomhrule\tableaustep
		\hbox{\phantomvrule\tableaustep\kern\tableaustep\phantomvrule\tableaustep}%
		\hbox{\vbox{\phantomhrule\tableauside}\kern-\tableaurule}}}
\def\squares#1{\hbox{\count0=#1\noindent\loop\sqr
		\advance\count0 by-1 \ifnum\count0>0\repeat}}
\def\tableau#1{\vcenter{\offinterlineskip
		\tableaustep=\tableauside\advance\tableaustep by-\tableaurule
		\kern\normallineskip\hbox
		{\kern\normallineskip\vbox
			{\gettableau#1 0 }%
			\kern\normallineskip\kern\tableaurule}%
		\kern\normallineskip\kern\tableaurule}}
\def\gettableau#1 {\ifnum#1=0\let\next=\null\else
	\squares{#1}\let\next=\gettableau\fi\next}

\newcommand{\be}{ \begin{equation}}
\newcommand{\ee}{\end{equation}}
\newcommand{\bea}[1]{\begin{eqnarray}\label{#1} }
\newcommand{\eea}{\end{eqnarray}}

\def\ZZZ{{\hskip-3pt\hbox{ Z\kern-1.6mm Z}}}
\def\zzz{{\hskip-3pt\hbox{ z\kern-1mm z}}}

\def\s{\sigma}
\def\bal#1\eal{\begin{align}#1\end{align}}

\def\one{{\hbox{ 1\kern-.8mm l}}}
\def\zero{{\hbox{ 0\kern-1.5mm 0}}}

\title{Higher Spins and Yangian Symmetries}

\author{Matthias R.\ Gaberdiel$^a$,  Rajesh Gopakumar$^b$, Wei Li$^c$  and Cheng Peng$^d$} 
\affiliation{$^a$ Institut f\"ur Theoretische Physik, ETH Zurich, \\
\hspace*{0.3cm}CH-8093 Zurich, Switzerland}
\affiliation{$^b$ International Centre for Theoretical Sciences-TIFR, \\
\hspace*{0.3cm}Survey No. 151, Shivakote, Hesaraghatta Hobli, \\
\hspace*{0.3cm}Bengaluru North, India 560 089}
\affiliation{$^c$ CAS Key Laboratory of Theoretical Physics, 
Institute of Theoretical Physics,\\ 
\hspace*{0.3cm}Chinese Academy of Sciences, 100190 Beijing, P.R.\ China}
\affiliation{$^d$ Department of Physics, Brown University, \\
\hspace*{0.3cm}182 Hope Street, Providence, RI 02912, USA}

\emailAdd{gaberdiel@itp.phys.ethz.ch, rajesh.gopakumar@icts.res.in, weili@itp.ac.cn, cheng$\underline{~}$peng@brown.edu}

\abstract{The relation between the bosonic higher spin ${\cal W}_\infty[\lambda]$ 
algebra, the affine Yangian of $\mathfrak{gl}_{1}$,
and the SH$^c$ algebra is established in detail. For generic $\lambda$ we find 
explicit expressions for 
the low-lying 
${\cal W}_\infty[\lambda]$ modes in terms of the affine Yangian generators, 
and deduce from this the
precise identification between $\lambda$ and the parameters of the affine Yangian.
Furthermore, for the free field cases corresponding to $\lambda=0$ and $\lambda=1$ 
we give closed-form expressions for the affine Yangian generators in terms of the 
free fields. Interestingly, the relation between the ${\cal W}_\infty$ modes
and those of the affine Yangian is a non-local one, in general. 
We also establish the explicit dictionary between the affine 
Yangian and the SH$^c$ generators. Given
that Yangian algebras are the hallmark of integrability, these identifications 
should pave the way towards uncovering the relation between the integrable 
and the higher spin 
symmetries.}

\begin{document}

\maketitle

\makeatletter
\g@addto@macro\bfseries{\boldmath}
\makeatother
\section{Introduction}

The higher spin -- CFT duality allows one to get a glimpse at the large symmetry algebra underlying
string theory \cite{Gross:1988ue,Witten:1988zd,Moore:1993qe}. Indeed, the higher spin symmetry of string
theory is believed to appear at the tensionless point in AdS \cite{Sundborg:2000wp,Witten,Mikhailov:2002bp}, where
infinitely many  fields of spin greater than two become massless and give rise to a Vasiliev higher spin theory 
\cite{Vasiliev:2003ev}. A concrete description of this phenomenon was found for the case of AdS$_3$ in \cite{Gaberdiel:2014cha},
and at least for this specific background, it is now possible to study the large unbroken symmetry algebra of string theory in detail.

The tensionless point of string theory on AdS is dual to the free limit of the dual field theory, at which the 
integrable structure of the field theory should have an explicit realization. It is then a natural question how the integrable structure relates to the higher spin and the stringy symmetries. In this paper we begin
to explore this question by studying the relation between the higher spin (and the stringy) symmetry algebra on AdS$_3$, and
certain Yangian algebras. Yangian algebras appear naturally in spin-chain models and are a 
hallmark of integrability. More specifically, we shall explore this question for the original bosonic version of the 
higher spin -- CFT duality \cite{Gaberdiel:2010pz}.
\smallskip

Some time ago it was noted  \cite{Tsymbaliuk,Prochazka:2015deb} that the affine 
Yangian of $\mathfrak{gl}_{1}$ (as defined in \cite{Tsymbaliuk,Tsymbaliuk14})
is isomorphic to ${\cal W}_{1+\infty}[\lambda]$, the asymptotic symmetry algebra of 
the bosonic higher spin theory on AdS$_3$.\footnote{Strictly speaking, the actual symmetry algebra is ${\cal W}_\infty[\lambda]$,
and does not contain the spin-one current, but the spin-one current can be easily decoupled and does not play an important
role in the following.} This isomorphism arises  as the rational limit of the 
equivalence between the quantum-deformed ${\cal W}_{1+\infty}$ algebra and the quantum toroidal algebra of $\mathfrak{gl}_1$  \cite{Tsymbaliuk}, 
generalizing the construction of \cite{GautamLaredo} to the toroidal case. The toroidal isomorphism was first pointed out in \cite{Miki}, and the definition of the 
quantum toroidal algebra (or quantum affinization of the affine Lie algebra) of $\mathfrak{gl}_1$ was inspired by 
\cite{Drinfeld}; the toroidal isomorphism was also independently re-derived in a series 
of papers  \cite{FFJMM1,FFJMM2,FJMM1}. More recently, the construction of the quantum toroidal
algebras was generalized further to arbitrary quiver diagrams in \cite{Kimura:2015rgi}.

In \cite{Prochazka:2015deb} Prochazka 
proposed a concrete dictionary for how the parameters of the affine 
Yangian of $\mathfrak{gl}_{1}$ and ${\cal W}_{1+\infty}[\lambda]$
are related to each other, and gave circumstantial
evidence for this by comparing the structure of some representations. In this paper we confirm his claim independently by
constructing the low-lying ${\cal W}_\infty[\lambda]$ generators explicitly in terms of the affine Yangian generators. This allows us to 
determine the two structure constants that characterize the ${\cal W}_\infty[\lambda]$ algebra \cite{Gaberdiel:2012ku}  --- the central 
charge $c$ and the OPE coefficient $C_{33}^{4}$ describing the coupling of two spin-$3$ fields to the spin-$4$ field --- 
and thereby check the proposed dictionary. We also establish this identification for the two free field cases, 
the free fermion case corresponding to $\lambda=0$ (which
 was already analysed in \cite{Prochazka:2015deb}) as well
as the the free boson construction leading to the algebra with $\lambda=1$.

One of the interesting lessons of our analysis is that in general the modes of the 
local fields of the ${\cal W}_\infty[\lambda]$ algebra involve
 infinite linear combinations of the Yangian generators, reflecting the 
inherently non-local structure of the Yangian algebra. Our analysis also allows us to clarify the way in which the triality symmetry
of the ${\cal W}_\infty[\lambda]$ algebra arises in the affine Yangian description. 
\smallskip

The affine Yangian of $\mathfrak{gl}_1$ is also believed to be 
isomorphic to the spherical degenerate double affine Hecke algebra, also called 
SH$^c$ algebra, of \cite{SV}. (In fact, the affine Yangian of $\mathfrak{gl}_1$ was first constructed (in the RTT formulation) by \cite{MO} at around the same time as  \cite{SV},  in the 
context of proving the AGT correspondence.)
While this relation seems to be known to the experts, an explicit description of the underlying
isomorphism does not seem to exist in the literature, and we have therefore also included an explicit account of it here. Among other
things this allows us to explain how different natural classes of representations are related to one another. 
\smallskip

All of these considerations concern the original higher spin symmetry algebra ${\cal W}_\infty$, rather than the stringy symmetry algebra,
i.e.\ the Higher Spin Square \cite{Gaberdiel:2015mra}. This therefore suggests that the integrable structure should already be visible in 
terms of the higher spin theory, and may not require including the stringy degrees of freedom. It would nevertheless be very interesting 
to understand how the Higher Spin Square algebra can be brought into the fold.
On the face of it, it also seems to correspond to some Yangian
algebra \cite{Gaberdiel:2015mra}, but this does not seem to be directly related to the Yangian algebras we encounter in this paper. 
\medskip

The paper is organized as follows. In Section~\ref{sec2} we briefly review the structure of the affine Yangian of 
$\mathfrak{gl}_{1}$ and the higher spin algebra ${\cal W}_\infty[\lambda]$. Section~\ref{sec:comp} is concerned
with working out the relation between the two structures. We also explain 
(in Sections~\ref{sec:triality} and \ref{sec:reps})
how the triality symmetry is realized, and how the representations can be identified. Section~\ref{sec:freefield} deals with the two 
free field realizations that provide specific incarnations of the general dictionary, and in Section~\ref{sec:SHc} we 
establish the detailed correspondence with the SH$^c$ algebra. Finally, Section~\ref{sec:conclusion} contains our conclusions.
There is one appendix in which some of the details of the construction of the spin-$4$ field of ${\cal W}_\infty[\lambda]$ in terms of affine Yangian
generators is explained in some detail.

\section{The affine Yangian and ${\cal W}_{1+\infty}[\lambda]$}\label{sec2}

We begin by reviewing the structure of the affine Yangian in Section~\ref{sec:AY}, and that of 
${\cal W}_{1+\infty}[\lambda]$ in Section~\ref{sec:Winf}.

\subsection{The affine Yangian of $\mathfrak{gl}_{1}$}\label{sec:AY}

The affine Yangian of $\mathfrak{gl}_{1}$ is the associative algebra generated by the generators
$e_j$, $f_j$, and $\psi_j$ with $j=0,1,\ldots$, subject to a set of commutation and anti-commutation relations
that are most easily described in terms of the generating functions \cite{Tsymbaliuk} 
\be\label{generating}
e(z) = \sum_{j=0}^{\infty} \, \frac{e_j}{z^{j+1}} \ , \qquad 
f(z) = \sum_{j=0}^{\infty} \, \frac{f_j}{z^{j+1}} \ , \qquad 
\psi(z)  = 1 + \sigma_3 \, \sum_{j=0}^{\infty} \frac{\psi_j}{z^{j+1}} \ , 
\ee
where $z$ is a `spectral' parameter, and $\sigma_3=h_1 h_2 h_3$. Here 
$(h_1,h_2,h_3)$ is a triplet of parameters whose sum is zero
\be
\sigma_1 \equiv h_1 + h_2 + h_3 = 0 \ , 
\ee
and we denote their symmetric powers by 
\be\label{sigma23}
\sigma_2 \equiv h_1 h_2 + h_2 h_3 + h_1 h_3 \ , \qquad \sigma_3 \equiv h_1 h_2 h_3 \ . 
\ee
In order to describe the relations we introduce the function 
\be\label{varphidef}
\varphi(z) = \frac{(z+h_1) (z+h_2) (z+h_3)}{(z-h_1) (z-h_2) (z-h_3)}  = \frac{z^3 + \sigma_2 z + \sigma_3}{z^3 + \sigma_2 z - \sigma_3}\ , 
\ee
which satisfies the obvious identity
\be
\varphi(z) \, \varphi(-z) = 1 \ .
\ee
The relations of the affine Yangian are then 
\begin{align}
e(z)\, e(w) \, \ \sim \ \ & \varphi(z-w)\, e(w)\, e(z)  \label{eegen} \\
f(z)\, f(w) \, \ \sim \ \ & \varphi(w-z)\, f(w)\, f(z) \label{ffgen} \\
\psi(z)\, e(w) \, \ \sim \ \ & \varphi(z-w)\, e(w)\, \psi(z) \label{psiegen} \\
\psi(z)\, f(w) \, \ \sim \ \ & \varphi(w-z)\, f(w)\, \psi(z) \ , \label{psifgen} 
\end{align}
where `$\sim$' means equality up to terms that are regular at $z=0$ or $w=0$. (Note that in
eq.~(\ref{generating}) the generators of the algebra are all associated to singular terms of the 
spectral parameter.) In addition we have the identity
\be\label{efgen}
e(z)\, f(w) - f(w)\, e(z) = - \frac{1}{\sigma_3}\, \frac{\psi(z) - \psi(w)}{z-w} \ ,
\ee
as well as the Serre relations 
\be
\sum_{\pi \in S_3}\, \bigl(z_{\pi(1)} - 2 z_{\pi(2)} + z_{\pi(3)} \bigr)\, e(z_{\pi(1)})\, e(z_{\pi(2)})\, e(z_{\pi(3)}) = 0 \ , 
\ee
together with the same identity for $f(z)$. This formulation is particularly suited 
for describing the 
representation theory of the affine Yangian  as we will see at the end of this section.
\medskip

In order to get a feeling for what these relations mean, it is useful to write them out in terms of modes. For 
example, multiplying the first equation (\ref{eegen}) by the denominator of (\ref{varphidef}) we obtain
\be
\sigma_3 \bigl\{ e(z) , e(w)  \bigr\}  = \Bigl( (z-w)^3 + \sigma_2 (z-w) \Bigr)\, \bigl[ e(z) ,e(w) \bigr]   \ ,
\ee
which upon expanding out in terms of modes leads to the relation 
\begin{eqnarray}
\sigma_3 \{e_j,e_k \}  & = & [e_{j+3},e_k] - 3  [e_{j+2},e_{k+1}] + 3  [e_{j+1},e_{k+2}]  - [e_{j},e_{k+3}]   \nonumber \\
& & \ + \sigma_2 [e_{j+1},e_{k}] - \sigma_2  [e_{j},e_{k+1}]  \ . \label{Y1} 
\end{eqnarray}
The other cases work similarly, and we find in addition the identities (see also \cite{Prochazka:2015deb})
\begin{eqnarray}
0 & = & [\psi_j,\psi_k] \label{Y0} \\
-  \sigma_3 \{f_j,f_k \} & = &  [f_{j+3},f_k] - 3  [f_{j+2},f_{k+1}] + 3  [f_{j+1},f_{k+2}]  - [f_{j},f_{k+3}]  \nonumber \\
& & \ + \sigma_2 [f_{j+1},f_{k}] - \sigma_2  [f_{j},f_{k+1}]   \label{Y2} \\
\psi_{j+k} & = & [e_j,f_k] \label{Y3} \\
\sigma_3 \{\psi_j,e_k \}  & = &  [\psi_{j+3},e_k] - 3  [\psi_{j+2},e_{k+1}] + 3  [\psi_{j+1},e_{k+2}]  - [\psi_{j},e_{k+3}]  \nonumber \\
& & \ + \sigma_2 [\psi_{j+1},e_{k}] - \sigma_2  [\psi_{j},e_{k+1}]  \label{Y4} \\
- \sigma_3 \{\psi_j,f_k \}  & = &  [\psi_{j+3},f_k] - 3  [\psi_{j+2},f_{k+1}] + 3  [\psi_{j+1},f_{k+2}]  - [\psi_{j},f_{k+3}]  \nonumber \\
& & \ + \sigma_2 [\psi_{j+1},f_{k}] - \sigma_2  [\psi_{j},f_{k+1}]  \label{Y5} \ , 
\end{eqnarray}
together with the low order relations
\be\label{ini1}
[\psi_0,e_j] = 0 \ , \qquad [\psi_1,e_j] = 0 \ , \qquad [\psi_2,e_j] = 2 e_j  \ ,
\ee
and
\be\label{ini2}
[\psi_0,f_j] = 0 \ , \qquad [\psi_1,f_j] = 0 \ , \qquad [\psi_2,f_j] = - 2 f_j  \ ,
\ee
as well as the Serre relations
\be\label{Serre}
{\rm Sym}_{(j_1,j_2,j_3)} [e_{j_1},[e_{j_2},e_{j_3+1}]] = 0 \ , \qquad
{\rm Sym}_{(j_1,j_2,j_3)} [f_{j_1},[f_{j_2},f_{j_3+1}]] = 0 \ . 
\ee
\medskip

It is immediate from the above commutation relations that the affine Yangian contains two central elements, namely
$\psi_0$ and $\psi_1$. In a given representation, the algebra is therefore characterized by four independent 
parameters: $\psi_0$ and $\psi_1$, together with $\sigma_2$ and $\sigma_3$. However, not all four parameters
are independent since the algebra possesses a scaling symmetry 
\be\label{psialpha}
\psi_j \mapsto \alpha^{j-2} \, \psi_j \ , \qquad e_j \mapsto \alpha^{j-1} e_j \ , \qquad f_j \mapsto \alpha^{j-1} f_j \ ,
\ee
under which the above relations remain invariant provided we also scale $\sigma_2 \mapsto \alpha^2 \sigma_2$ and
$\sigma_3 \mapsto \alpha^3 \sigma_3$, i.e. 
\be\label{halpha}
h_j \mapsto \alpha \, h_j \ . 
\ee
Thus the algebra actually only depends on three  of the four parameters $\sigma_2$, $\sigma_3$, $\psi_0$ and $\psi_1$;
in particular, we may consider the scale-invariant combinations
\be\label{scaleinv}
\sigma_2 \psi_0 \ , \qquad  \sigma_3^2\, \psi_0^3 \ , \qquad \hbox{and} \qquad  \psi_1^2 \psi_0^{-1} \ . 
\ee

\begin{figure}[!h]
	\centering
	\begin{tabular}{c}
		\includegraphics[width=.22\textwidth]{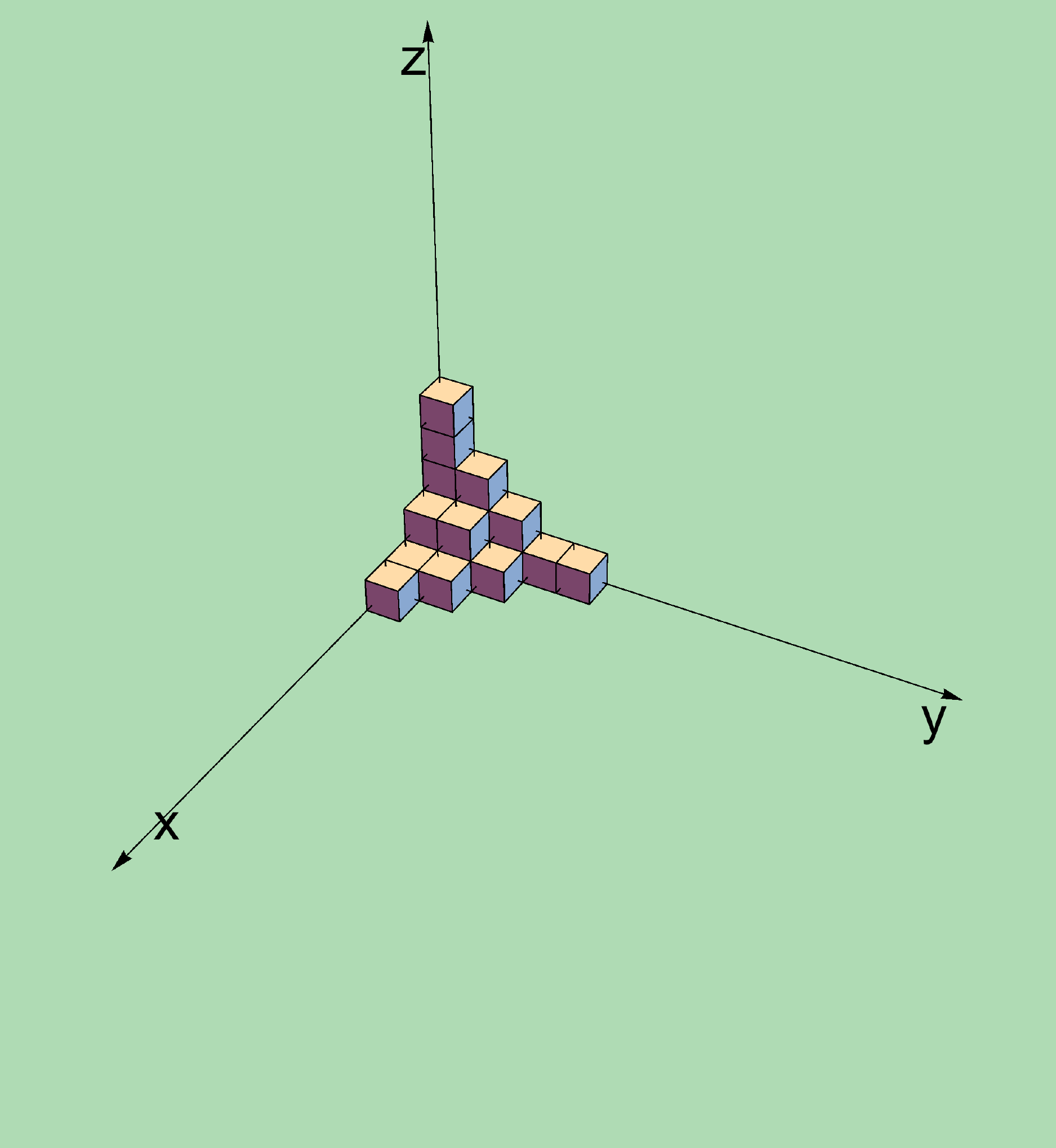} \, 
		\includegraphics[width=.22\textwidth]{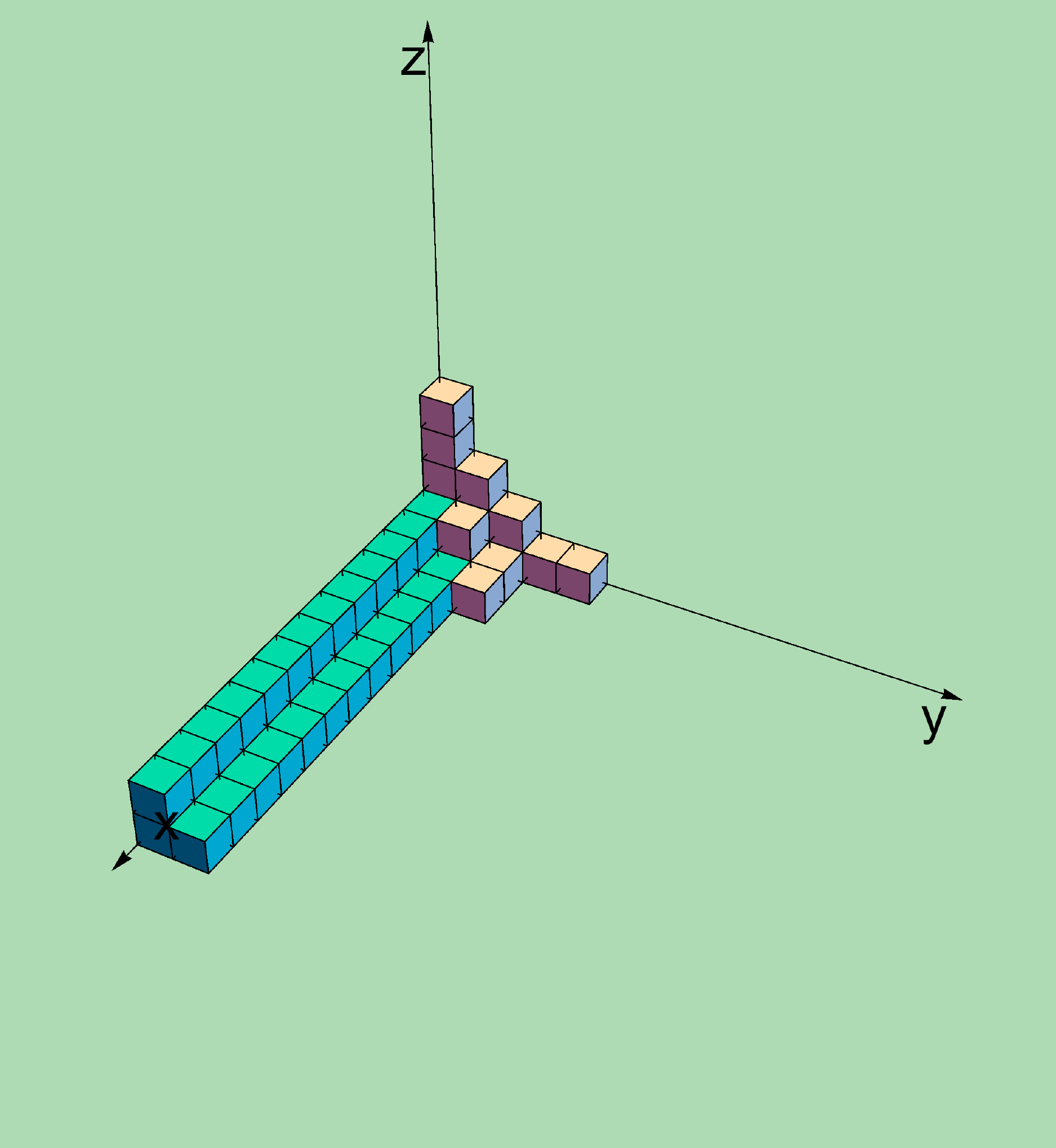}\,
		\includegraphics[width=.22\textwidth]{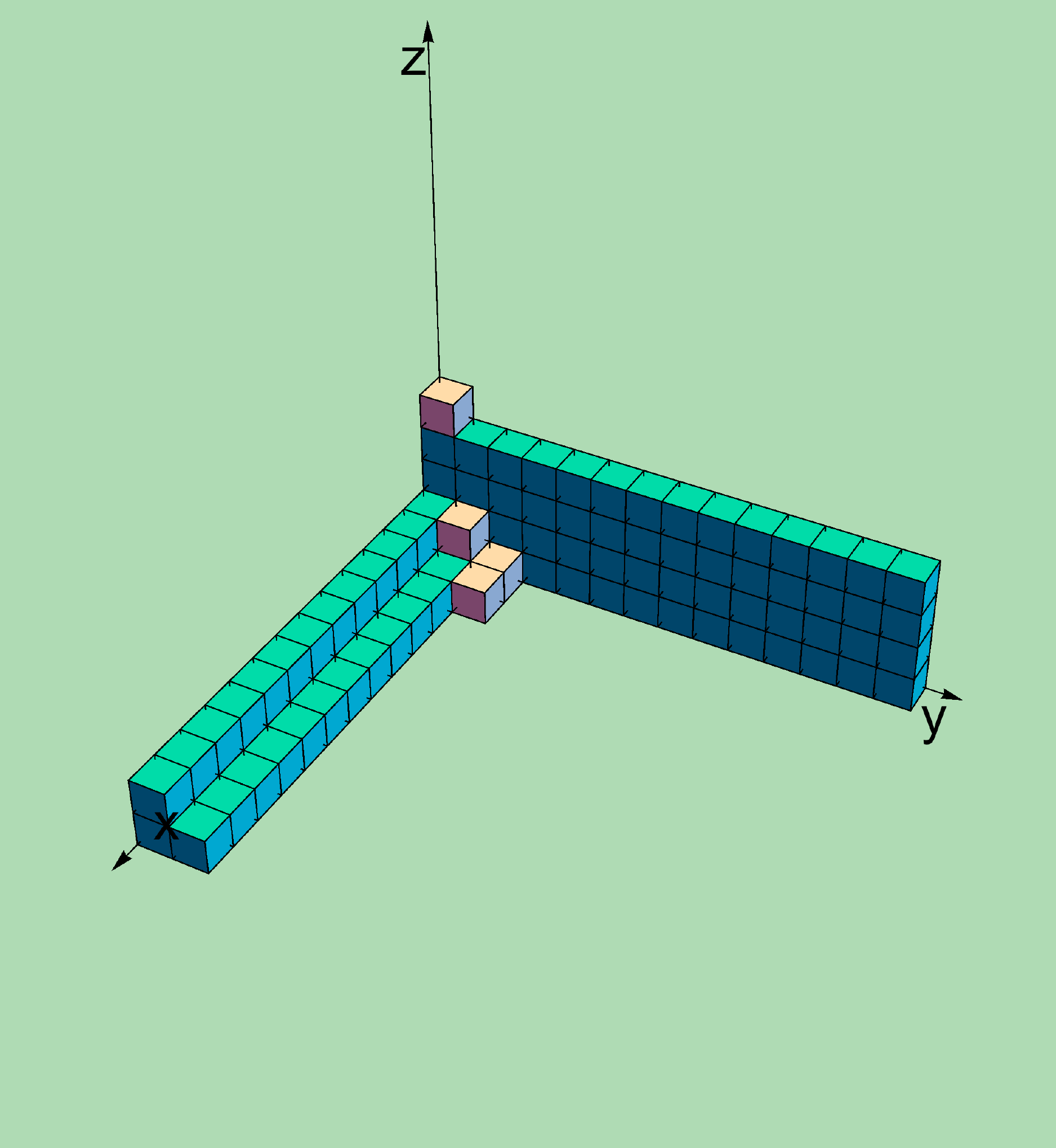} \, 
		\includegraphics[width=.22\textwidth]{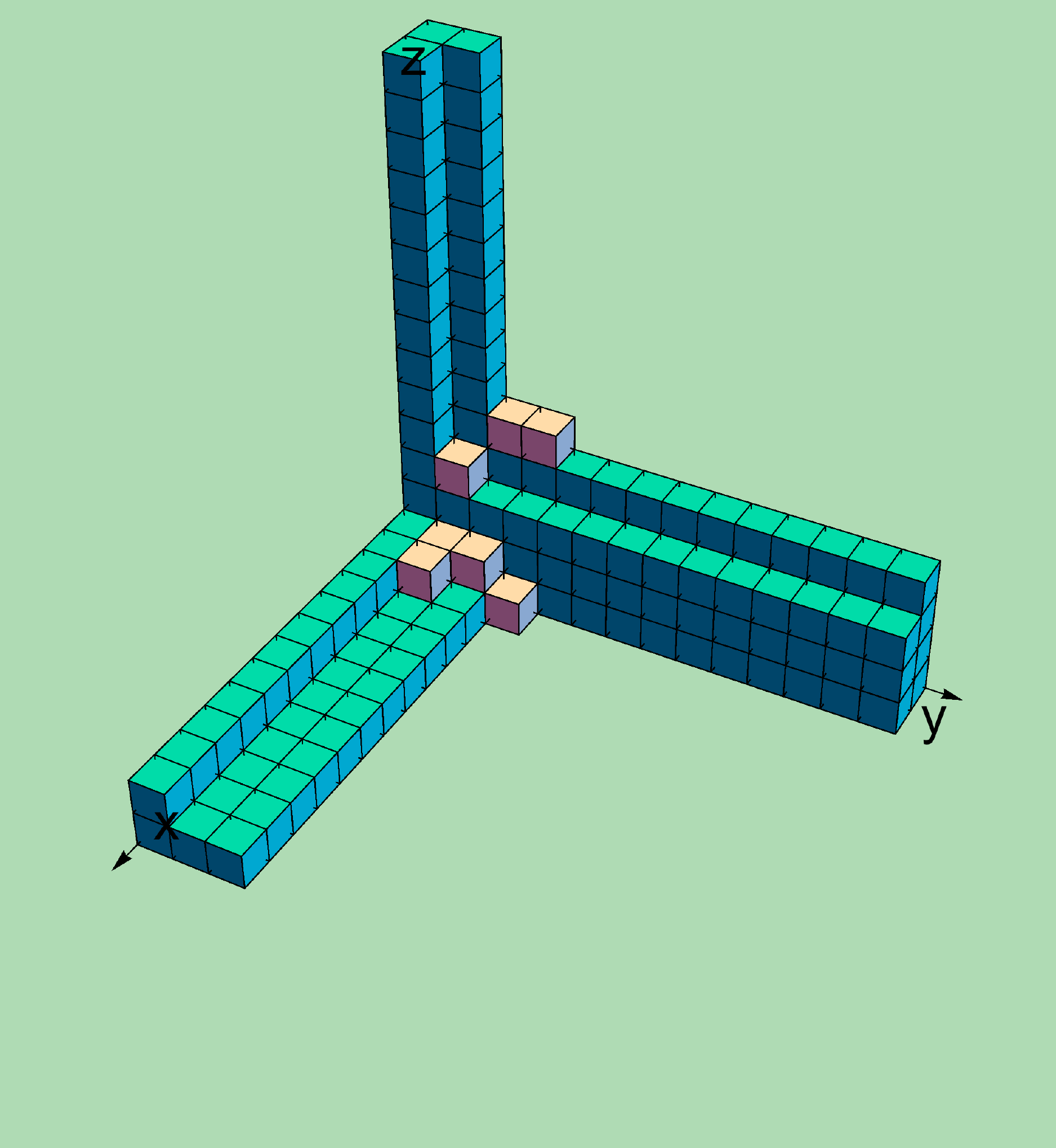}   \\
		\vspace{-1.0cm} \qquad vacuum \qquad\qquad \quad $(\Xi,0)$  \qquad\qquad\qquad   $(\Xi_{+},\Xi_{-})$ 
		  \qquad\quad $(\Xi^{(x)} ,\Xi^{(y)} ,\Xi^{(z)} )$ \vspace{1.5cm}
	\end{tabular}
	\caption{There are four types of plane partition representations, depending on how many of the three asymptotics are non-trivial.}
	\label{fig1}
\end{figure}

There is a very natural class of representations of the affine Yangian of $\mathfrak{gl}_1$ which are of interest in the 
connection to the ${\cal W}_{\infty}$ algebra.\footnote{Strictly speaking,
they define representations of ${\cal W}_{1+\infty}$, but as explained below,
see Section~\ref{sec:Winf}, the $\mathfrak{u}(1)$ part can be easily decoupled.}
These representations are best viewed in terms of  plane partitions, i.e.\ 
three-dimensional box stacking configurations.\footnote{For a good introduction
on combinatorics of plane partitions, see \cite{Stanley}.} They
are special in that they have a finite number of states at every level and thus 
possess infinitely many null states; they are 
discussed in some detail in \cite{Prochazka:2015deb,Datta:2016cmw}, 
and we only summarize the salient aspects which we will draw upon later. 
A generic plane partition representation is labelled by three Young tableaux. These Young tableaux correspond to the asymptotic box 
configurations (along the three positive axes) of a three dimensional box stacking configuration. A valid stacking is one in which the number of 
boxes are non-increasing as one moves from any site to a neighboring site along any of the three positive directions (see Figure~\ref{fig1}).

 \begin{comment}
 \begin{figure}[!h]
 	\centering
 	\begin{tabular}{c}
 		\includegraphics[width=.3\textwidth]{PPtype0.pdf} \qquad \qquad 
 		\includegraphics[width=.3\textwidth]{PPtype1.pdf}  \\ 
 			\includegraphics[width=.3\textwidth]{PPtype2.pdf} \qquad \qquad 
 		\includegraphics[width=.3\textwidth]{PPtype3.pdf}  \\ 
 		\vspace{-1.5cm} (a) \hspace{5.5cm} (b) \vspace{1.5cm}
 	\end{tabular}
 	\caption{Depending on how many of the three asymptotics is non-trivial, there are four types of plane partition representation.}
 	\label{fig3}
 \end{figure}
\end{comment}

The highest weight state of a representation (labelled by three Young tableaux) is the 
plane partition configuration with the minimum number of 
boxes consistent with the specified asymptotics (the blue boxes in Figure~\ref{fig1}).
The other states in the representation are 
obtained by the repeated action of the Yangian generators on this representation (given by the yellow boxes in Figure~\ref{fig1}). 
This action is given in terms of adding/removing boxes from
a given valid stacking configuration. To be specific, we have the action of the generating functions (\ref{generating}) \cite{Prochazka:2015deb}
\begin{eqnarray}\label{ppart}
\psi(z)|\Lambda \rangle & = &\psi_{\Lambda}(z)|\Lambda \rangle  \ ,\label{psizdef}\\
e(z) | \Lambda \rangle & = & \sum_{ {\tiny \yng(1)} \in {\rm Add}(\Lambda)}
\frac{\bigl[ -  \frac{1}{\sigma_3} {\rm Res}_{w= h({\tiny \yng(1)})} \psi_{\Lambda}(w) \bigr]^{\frac{1}{2}}}{ z - h({\tiny \yng(1)}) } 
| \Lambda + {\tiny \yng(1)} \rangle \ , \label{ezdef} \\
f(z) | \Lambda \rangle & = & \sum_{ {\tiny \yng(1)} \in {\rm Rem}(\Lambda)}
\frac{\bigl[ -  \frac{1}{\sigma_3} {\rm Res}_{w= h({\tiny \yng(1)})} \psi_{\Lambda}(w) \bigr]^{\frac{1}{2}}}{ z - h({\tiny \yng(1)}) } 
| \Lambda - {\tiny \yng(1)} \rangle \ , \label{fzdef}
\end{eqnarray}
where `Res' denotes the residue. 
Thus $\psi(z)$ acts diagonally on a plane partition 
configuration $\Lambda$ with eigenvalue 
\be
\psi_\Lambda(z) = \bigl( 1 + \frac{\psi_0 \sigma_3}{z} \bigr) \, \prod_{ {\tiny \yng(1)} \in \Lambda} \varphi(z - h({\tiny \yng(1)}) ) \ , 
\ee
where 
\be\label{hbox}
h({\tiny \yng(1)}) = h_1 x({\tiny \yng(1)}) + h_2 y({\tiny \yng(1)}) +  h_3 z({\tiny \yng(1)}) 
\ee
with $x({\tiny \yng(1)})$ the $x$-coordinate of the box, etc. The RHS of the second and third lines indicate that the action of $e(z)$ and 
$f(z)$ give rise to configurations with one more (or one less) box. The sum is over all 
plane partition configurations of this kind.

 In this language, the vacuum representation of ${\cal W}_{1+\infty}$ is the plane partition with trivial asymptotics, i.e., its highest 
 weight configuration has no boxes. The character of this representation thus counts all finite plane partitions 
 whose generating function is given by the MacMahon function. The minimal representation is the next simplest case, which has a single box Young tableau boundary condition in one of the three directions (and trivial asymptotic behaviour in the other two). 
 We immediately see that triality, which acts by interchanging the $h_i$ and  the three axes, gives rise to two other minimal 
 representations, which have the same character (up to an overall factor involving the conformal dimension of the highest weight), in 
 agreement with the prediction of \cite{Gaberdiel:2012ku}.

\subsection{The algebra ${\cal W}_{1+\infty}[\lambda]$}\label{sec:Winf}

The algebra ${\cal W}_{1+\infty}[\lambda]$ is the ${\cal W}$ algebra 
that is generated by one independent field for each spin $s=1,2,3,\ldots$. Since
it contains a spin-one field, whose commutator can only contain a central term and hence define a $\mathfrak{u}(1)$ algebra, 
\be\label{JJ}
[J_m, J_n ] = m \, \kappa\, \delta_{m,-n} \ , 
\ee
we can consider the coset of the ${\cal W}_{1+\infty}[\lambda]$ algebra by this $\mathfrak{u}(1)$ field, and hence conclude that 
\be\label{u1dec}
{\cal W}_{1+\infty}[\lambda] \, \cong \, \mathfrak{u}(1) \ \oplus \  {\cal W}_{\infty}[\lambda] \ , 
\ee
see, e.g., the discussion in \cite{Gaberdiel:2013jpa}. Here ${\cal W}_\infty[\lambda]$ is the ${\cal W}$ algebra generated by one field for each 
spin $s=2,3,\ldots$; as was shown in \cite{Gaberdiel:2012ku}, this algebra is uniquely characterized by the value of 
the central charge $c$ and the OPE coupling constant $C^4_{33}$ describing the 
coupling of two spin-$3$ fields to the spin-$4$ field.
This OPE coefficient is parametrized in terms of $c$ and $\lambda$ as 
\be \label{c433}
(C^4_{33})^2  =  
\frac{64 (c+2) (\lambda-3) \bigl( c(\lambda+3) + 2 (4\lambda+3)(\lambda-1) \bigr)}{
(5c+22) (\lambda-2)  \bigl( c(\lambda+2) + (3\lambda+2)(\lambda-1)\bigr)} \ .
\ee
Since the numerator and denominator of (\ref{c433}) are cubic polynomials in $\lambda$, there are generically three different values of 
$\lambda$ (which also depend on $c$ since the coefficients 
of the polynomials are functions of $c$)  that lead to the same OPE structure 
constant $C^4_{33}$ and therefore to the same algebra. This is referred to as the `triality' symmetry of the algebra. 

In addition to the two parameters $(\lambda,c)$, the algebra ${\cal W}_{1+\infty}[\lambda]$ is also characterized by the 
central term $\kappa$ in (\ref{JJ}), as well as the eigenvalue of the central generator $J_0$. (The coset construction guarantees
that $J_0$ commutes with all generators of ${\cal W}_\infty[\lambda]$.) 
Obviously, $\kappa$ can be rescaled by rescaling the $J_m$
generators, but this would also modify the value of $J_0$; 
therefore the (scale) invariant quantity is 
\be\label{Jpar}
(J_0)^2 / \kappa   \ . 
\ee
Hence the ${\cal W}_{1+\infty}[\lambda]$ algebra is also characterized by three parameters, which we may take to be 
$\lambda$, $c$, and the ratio (\ref{Jpar}).

\section{Relating ${\cal W}_{1+\infty}[\lambda]$ and the 
affine Yangian of $\mathfrak{gl}_1$}\label{sec:comp}

In this section we explain in detail the relation between ${\cal W}_{1+\infty}[\lambda]$ and the affine Yangian of 
$\mathfrak{gl}_{1}$. We start by identifying the spin-one and spin-two generators.

\subsection{Identification of generators}

\subsubsection{The spin-one and spin-two generators} 

Following \cite{Prochazka:2015deb} we identify the $\mathfrak{u}(1)$ algebra of ${\cal W}_{1+\infty}[\lambda]$ 
inside the affine Yangian by setting 
\be\label{u1gen}
J_1=-f_0\ , \qquad J_{-1}=e_0\ , 
\ee
so that 
\be
\kappa =  \psi_0 \ . 
\ee
The higher modes of the $\mathfrak{u}(1)$ algebra can be obtained recursively by using the commutation relations 
\be\label{LJ}
{}[L_m,J_n ] = -n J_{m+n} \ . 
\ee
(Note that these $L_m$ generators are the full Virasoro generators of the ${\cal W}_{1+\infty}[\lambda]$ algebra;
in particular, they do not commute with the $\mathfrak{u}(1)$ generators.) In fact, it is enough to use the commutation relations
with the M\"obius generators $L_{\pm 1}$ (instead of all Virasoro generators $L_m$), for which we make the ansatz 
\be\label{Virgen}
L_1 =-f_1\ , \qquad L_{-1}=e_1\ .
\ee
Then the scaling operator $L_0$ is
\be\label{L0gen}
L_0 = \tfrac{1}{2}\, [L_1,L_{-1}]=\tfrac{1}{2}\, \psi_2\ ,
\ee
from which we deduce that $J_0$ equals
\be\label{J0gen}
J_0 = [L_1, J_{-1}] =  -[f_1, e_0] =  \psi_1\ . 
\ee
Note that (\ref{J0gen}) is then also equal to 
$\psi_1 =   [e_1, f_0] = - [L_{-1}, J_1]$, as
required by consistency. 
The scale-invariant combination from (\ref{scaleinv})
\be
\psi_1^2 \psi_0^{-1} \cong  (J_0)^2 / \kappa \ , 
\ee
matches then precisely with the invariant combination (\ref{Jpar}).

In order to identify the higher Virasoro generators, we need to specify the form of the generators $L_{\pm 2}$, for which we 
make the ansatz, again following \cite{Prochazka:2015deb}
\be
L_{-2}=\tfrac{1}{2}\bigl([e_2,e_0]+\sigma_3 \psi_0\, [e_0,e_1]\bigr)\ , \qquad L_{2}=-\tfrac{1}{2}\bigl([f_2,f_0]+\sigma_3\psi_0\, [f_0,f_1]\bigr)\ . 
\ee
Using the commutation relations of the Virasoro algebra, this then determines recursively all $L_m$ generators (upon
taking repeated commutators with $L_1$ and $L_{-1}$, respectively). The resulting generators then satisfy --- we have only
checked this explicitly for the first few cases --- 
\be\label{cval}
[L_m,L_n] = (m-n) L_{m+n} + \frac{c}{12} m (m^2-1) \, \delta_{m,-n} \ , \qquad \hbox{with} \quad
c = - \sigma_2 \psi_0\, - \sigma_3^2 \psi_0^3  \ . 
\ee
Note that the parameters that appear in the definition of $c$ are indeed scale-invariant, see eq.~(\ref{scaleinv}). In addition,
the resulting generators are compatible with eqs.~(\ref{LJ}) and (\ref{JJ}).

\subsubsection{The spin-three generators}

For the wedge modes of the spin-three generator we now make the ansatz 
\begin{align}
W^3_{-2}&=-[e_1,e_2] \\
W^3_{-1}&=-e_2-\tfrac{1}{2}\sigma_3  \psi_0\, e_1 \label{W31}\\
W^3_0&=\tfrac{1}{6}(-2\psi_3-\sigma_3 \psi_0 \psi_2 )\\
W^3_{1}&=f_2+\tfrac{1}{2}\sigma_3 \psi_0\, f_1 \label{W31m}\\
W^3_{2}&=-[f_1,f_2]\ .
\end{align}
These modes are constructed so as to have the correct commutation relations with the M\"obius generators
\be
[L_m, W^3_n]=(2m-n) W^3_{m+n}\qquad \hbox{for $m=0,\pm 1$} \ . \label{Wquasi}
\ee
However, the above ansatz cannot be quite correct since,
for $|n|\leq 2$, the commutators with the spin-one modes $J_m$ are  
\be
[J_{m},W^3_n]=-2 m L_{m+n}-\text{sign}(m)\,m(m+\tfrac{n}{2})\sigma_3\psi_0 J_{m+n}\ ,\label{gen13} 
\ee
and hence are not of the general form predicted by \cite{Blumenhagen:1990jv} based on the locality of the fields --- 
the offending term is the ${\rm sign}(m)$ term that is proportional to $\sigma_3$; 
it necessarily appears in the above
commutators as one can conclude, for example, from 
\begin{align}
[J_{-1},W^3_1]&=2 L_0+\tfrac{1}{2}\sigma_3 \psi_0 \, J_0 \\ 
[J_{1},W^3_2]&=-2 L_{3}-2\sigma_3\psi_0 \, J_{3} \ . 
\end{align} 
Thus the above $W^3_n$ modes cannot be the modes of a local field.
In order to repair this, we define the (non-local) generators (for $n\in\mathbb{Z}$)
\be
\tilde{W}_n  = \frac{1}{2} \sum_l |l - \tfrac{n}{2}|\, \Theta( l (l-n) )\, :J_{n-l} J_l : + \frac{1}{12} (|n|+2)(|n|+1) \, J_n J_0 \ , 
\ee
where $\Theta(m)$ is the step function defined by $\Theta(m)=0$ for $m\leq 0$, 
and $\Theta(m)=1$ for $m>0$. (Note that the $\Theta$-function term
guarantees that the mode numbers of the two $J$-modes in the first sum have opposite signs.)
The $\tilde{W}_n$ generators have the property that, for $m=0, \pm 1$ 
\be
{}[L_m,\tilde{W}_n] = (2m -n ) \tilde{W}_{m+n} \ . \label{tildeWquasi}
\ee
Furthermore, for $|n|\leq 2$, 
\be
{}[J_m,\tilde{W}_n] = {\rm sign} (m)\, m (m+\tfrac{n}{2})\,\psi_0\, J_{m+n} \ ,  \label{JtW}
\ee
which follows from the commutation relations  (\ref{JJ}). Thus we can modify the definition of $W^3_n$ 
by this non-linear correction term to remove the strange term in (\ref{gen13}). More specifically, if we define
\be
V^3_n=W^3_n + \sigma_3\, \tilde{W}^3_n\ ,\label{defV}
\ee
the redefined modes satisfy
\be
[J_{m},V^3_n]=-2 m L_{m+n}\ . \label{JV3}
\ee
This relation in fact remains true for the modes $W^3_n$ with $|n|\geq 3$, provided we define the
outside-the-wedge generators appropriately; a choice that  works is 
\be
W^3_3 =-\tfrac{1}{6}[f_2,[f_0,f_2]]+\tfrac{\sigma_3 \psi_0}{8}[f_2,[f_0,f_1]]-\tfrac{1}{12}(2\sigma_2+\sigma_3^2\psi_0^2
+\sigma_3 \psi_1)[f_1,[f_0,f_1]]+\tfrac{ \sigma_3}{4} [f_1,f_0^2]\ , \label{aztw33}
\ee
and similarly for $W^3_{-3}$, and then determining the remaining outside-the-wedge generators by recursively
considering the commutators with the M\"obius generators $L_{\pm 1}$.
Then we can calculate the commutators with the $J_m$ modes, and find for example
\begin{align}
[J_{-1},W^3_3]&=2L_2 \label{jm1w33}\\
[J_{-2},W^3_3]&=4L_1 \label{jm2w33}\\
[J_{-3},W^3_3]&=6L_0+5\sigma_3\psi_0 \, J_0 \label{jm3w33}\\
[J_{1},W^3_3]&=-2L_4-\tfrac{5}{2}\sigma_3\psi_0 \, J_4\,. \label{j1w33}
\end{align}
One can check that the terms proportional to the $J$-modes are again cancelled by the commutator with the correction term 
$\tilde{W}^3_3$, so that (\ref{JV3}) is indeed true for $n=3$ 
and $m=-1,-2,-3,1$. Since all the other commutators
can be recursively determined from these using the quasi-primary condition, i.e.\ the fact that  the $W^3$ and $\tilde{W}^3$ modes
satisfy (\ref{Wquasi}) and (\ref{tildeWquasi}), it follows that (\ref{JV3}) holds for all modes.

The $V^3_n$ modes are the modes of a quasi-primary, but not primary, spin-three field.
Indeed, we find for the commutators with the Virasoro generators
\be
[L_m,V^3_n] =(2m-n)V^3_{m+n}+\tfrac{m(m^2-1)}{6} \, (\s_2+\s_3^2\psi_0^2) \,  J_{m+n}\ .\label{LV3}
\ee
While this commutator still contains a correction term proportional to $J$, the structure of (\ref{LV3}) is now 
compatible with locality \cite{Blumenhagen:1990jv}.

\subsubsection{The spin-$4$ generators and the spin-$3$ commutators}

The analysis for the spin-$4$ generators is similar. 
Since the details are somewhat tedious, we have moved
the description of the construction to Appendix~\ref{app:Spin4}. With the definition of the spin-$4$ generators
at hand, we can then analyse the relevant commutators. First we determine by direct calculation the 
commutators of the $W^3$ modes 
\begin{align}
[W^3_0,W^3_2]&=-4 W^4_2 \label{w3w3}\\
[W^3_0,W^3_1]&=-2W^4_1+\frac{1}{10} (- \sigma_3^2 \psi_0^2+4  \sigma_3 \psi_1 -4 {\sigma_2 })L_1 \label{w3w3p} \ ,
\end{align}
where the $W^4$ modes are defined in eqs.~(\ref{W42}) and (\ref{W41}), respectively. 
Taking into account the correction terms, one then finds for the commutators of the $V^3$ modes 
\begin{align}
[V^3_0,V^3_2]&=-4 V^4_2 + \s_3^2\, \psi_0 \sum_{m\leq 0} (2m-2)^2 J_m J_{2-m} \label{V3V3}\\
[V^3_0,V^3_1]&=-2V^4_1+\frac{1}{10} (- \sigma_3^2\psi_0^2 -4 {\sigma_2 })L_1 +\frac{1}{2} \, \s_3^2\,  \psi_0
\sum_{m\leq 0} (2 m-1)^2 J_m J_{1-m} \label{V3V3p}\ ,
\end{align}
where the $V^4$ modes are defined in eq.~(\ref{V4def}). These commutators are now of local form --- the 
$JJ$ bilinear term has spin $s=4$.

\subsection{Decoupling the $\mathfrak{u}(1)$ currents}

The generators $V^3_n$ and $V^4_n$ are the modes of a local spin-$3$ and spin-$4$ field of ${\cal W}_{1+\infty}[\lambda]$,
respectively. However, these fields are neither $\mathfrak{u}(1)$ nor Virasoro primary, see, e.g.\ eqs.~(\ref{JV3}) and
(\ref{LV3}), and eqs.~(\ref{JV4}) and (\ref{LV4}), respectively. As a consequence, it is difficult to read off from their commutators
the relevant structure constants directly. 
On the other hand, we know on general grounds that we can 
decouple the $\mathfrak{u}(1)$ current, see eq.~(\ref{u1dec}), and that the resulting algebra must then be isomorphic to 
${\cal W}_\infty[\lambda]$. 
Thus it remains to perform this $\mathfrak{u}(1)$ decoupling explicitly, 
following \cite{Gaberdiel:2013jpa}. For the Virasoro generators,
the analysis is quite standard, and we find for the decoupled generators 
\be
\tilde{L}_m=L_m - \frac{1}{2\psi_0} \sum_l :J_l J_{m-l}:\ .
\ee
These generators then give rise to a Virasoro algebra 
\be
[\tilde{L}_m,\tilde{L}_n]
=(m-n)\tilde{L}_{m+n}+\frac{\tilde{c}}{12}\, m\, (m^2-1)\, \delta_{m,-n}\ , \qquad [\tilde{L}_m,J_n]=0\ , 
\ee
where the central charge equals now, cf.\ eq.~(\ref{cval}) 
\be\label{tildec}
\tilde{c} = -\sigma_2 \psi_0\, - \sigma_3^2 \psi_0^3 - 1 \ . 
\ee
For the $\mathfrak{u}(1)$ decoupled spin-$3$ field we find 
\be
\tilde{V}^3_m
=V^3_m + \frac{2}{\psi_0}\, \sum_l :J_{m-l}\tilde{L}_l:+ \frac{1}{3\psi_0^2}\, \sum_{l,n} :J_{m-n-l}J_nJ_l:\ .
\ee
This field is then primary with respect to both the $\mathfrak{u}(1)$ and the decoupled Virasoro algebra, i.e.
\be
[\tilde{L}_m,\tilde{V}^3_n] =(2m-n) \tilde{V}^3_{m+n}\ , \qquad [\tilde{V}^3_m,J_n]=0\ .
\ee
The $\mathfrak{u}(1)$ decoupled primary spin-$4$ wedge generators are 
\begin{align}
\nonumber V^4_m&=\tilde{V}^4_m-\frac{3}{\psi_0}\, \sum_l :\tilde{V}^3_{m-l}J_l:
+ \frac{3}{\psi_0^2}\, \sum_{n,l}:\tilde{L}_{m-n-l}J_n J_l:\\
\nonumber &\quad + \frac{1}{4\psi_0^3}\, 
\sum_{l,p,q} :J_{m-l-p-q}J_l J_p J_q:
- \frac{\s_2 -\s_3^2\psi_0^2} {20 \psi_0}\, \sum_{l}(5l^2-5ml+m^2+1) :J_{m-l}J_l:\\
&\quad +\, \frac{3(3\s_3^2\psi_0^3+7\s_2\psi_0+5) }{\psi_0(5\s_3^2\psi_0^3+5\s_2\psi_0-17)}
\Big(\sum_{l} :\tilde{L}_{m-l}\tilde{L}_l:-\frac{3}{10} (m+2)(m+3)\tilde{L}_m\Big)\ , \qquad 
\end{align}
which satisfy for $|n|\leq 3$ 
\be
[\tilde{L}_m,\tilde{V}^4_n]=(3m-n)\tilde{V}^4_{m+n}\ ,\qquad [J_m, \tilde{V}^4_n]=0\ . 
\ee

\subsection{Determining the structure constant}\label{sec:sc}

With these explicit expressions at hand, we can now determine the 
commutators of these $\mathfrak{u}(1)$ decoupled modes and find 
\begin{equation}
\begin{aligned}
{}[\tilde{V}^3_0,\tilde{V}^3_2]&=-4 \tilde{V}^4_2 +\frac{16(-\s_3^2\psi_0^3-4\s_2\psi_0-8) }{\psi_0(5\s_3^2\psi_0^3+5\s_2\psi_0-17)} 
\sum_{m=-\infty}^{\infty} :\tilde{L}_m \tilde{L}_{2-m}: \\
[\tilde{V}^3_0,\tilde{V}^3_1]
&=-2\tilde{V}^4_1+\frac{(-\s_3^2\psi_0^3-4\s_2\psi_0-8) }{10\psi_0} \tilde{L}_1  
\\
&\qquad +\frac{8(-\s_3^2\psi_0^3-4\s_2\psi_0-8) }{\psi_0(5\s_3^2\psi_0^3+5\s_2\psi_0-17)} 
\Big(\sum_{m=-\infty}^{\infty}:\tilde{L}_m\tilde{L}_{1-m}:+\tfrac{2}{5} \tilde{L}_1\Big)\ . 
\end{aligned}
\end{equation}
This can now be compared with eq.~(A.6) of \cite{Gaberdiel:2012ku}. Comparing \eqref{V3V3} to the 
case $m=0$, $n=2$  of (A.6) in \cite{Gaberdiel:2012ku}, we have the identifications
\begin{align}
\tilde{c} &=-\psi_0(\s_2+\s_3^2\psi_0^2)-1 \label{c1}\\
N_3&=\frac{\tilde{c}+\frac{22}{5}}{16}\frac{16(\s_3^2\psi_0^3+4\s_2\psi_0+8) }{\psi_0(5\s_3^2\psi_0^3+5\s_2\psi_0-17)}
= \frac{1}{5}\frac{(-\s_3^2\psi_0^2-4\s_2\psi_0-8) }{\psi_0 } \ . \label{N3}
\end{align}
In order to determine $\lambda$ it remains to compute at least one term in the commutator 
$[\tilde{V}^3_n,\tilde{V}^4_m]$; the simplest case to consider is the coefficient of the $\Theta^6_5$ term 
in the commutator $[\tilde{V}^3_2,\tilde{V}^4_3]$, where $\Theta^6$ is the $\mathfrak{u}(1)$ and 
Virasoro primary composite field of spin $6$, whose modes are of the form 
\be
\Theta^6_n = \sum_m\, (\tfrac{5}{3}m-n) : \tilde{L}_{n-m} \tilde{V}^3_{m}: + \ \hbox{terms proportional to $\tilde{V}^3$}\ ,
\ee
see eq.~(A.10) of \cite{Gaberdiel:2012ku}. From the explicit form of the commutators we can read off
the coefficient of $\Theta^6_5$ in the commutator $[\tilde{V}^3_2,\tilde{V}^4_3]$ to be 
\be
\frac{18(-27-9\s_2\psi_0-\s_3^2\psi_0^3)}{5\psi_0 (17-5\s_2\psi_0-5\s_3^2\psi_0^3)}\ .
\ee
Comparing to eq.~(3.4) of \cite{Gaberdiel:2012ku}, this coefficient should equal $\frac{84 N_4}{25 N_3 (c+2)}$ --- recall
that the central charge $c$ of \cite{Gaberdiel:2012ku} needs to be identified with $\tilde{c}$ defined by eq.~(\ref{c1}) 
--- from which we conclude that 
\be
\frac{N_4}{ N_3} =  \frac{15  (c+2)(-27-9\s_2\psi_0-\s_3^2\psi_0^3)}{14 \psi_0 (17-5\s_2\psi_0-5\s_3^2\psi_0^3)} \ . 
\ee
Together with the previous expression \eqref{N3}, we therefore find
\be
\frac{N_4}{N_3^2} = \frac{75  (c+2)(27+9\s_2\psi_0+\s_3^2\psi_0^3)}{14 (8+4\s_2\psi_0+\s_3^2\psi_0^3)(17-5\s_2\psi_0-5\s_3^2\psi_0^3)}\ .
\ee
This is now to be equated with, see eq.~(3.4) of \cite{Gaberdiel:2012ku},
\be
\frac{N_4}{N_3^2}= \frac{75 (c+2) ( \lambda-3) \bigl(c (\lambda+3) + 2 ( 4\lambda + 3)(\lambda-1)\bigr)}{14 (5c+22) (\lambda-2) 
\bigl(c (\lambda+2) + (3\lambda+2) (\lambda-1)\bigr)} \ , 
\ee
from which we conclude that 
\be\label{1.124}
\frac{(27+9\s_2\psi_0+\s_3^2\psi_0^3) }{(8+4\s_2\psi_0+\s_3^2\psi_0^3)  }
= \frac{( \lambda-3) \bigl(c (\lambda+3) + 2 ( 4\lambda + 3)(\lambda-1) \bigr)}{(\lambda-2) \bigl(c (\lambda+2) + (3\lambda+2) (\lambda-1)\bigr)} \ . 
\ee
In order to explain the relation between the two sets of parameters, it is now convenient to parameterize 
$c$ and $\lambda$ in terms of the coset labels $(N,k)$ via
\be
c = (N-1) \Bigl( 1 - \frac{N(N+1)}{(N+k)(N+k+1)} \Bigr) \ , \qquad 
\lambda = \frac{N}{N+k} \ ,
\ee
and the $\sigma_2$ and $\sigma_3$ variables in terms of $h_1$, $h_2$ and $h_3$, see eq.~(\ref{sigma23}). Then 
one checks by direct computation that the relations $c=\tilde{c}$ of eq.~(\ref{tildec}) and 
eq.~(\ref{1.124}) are exactly compatible with the proposed identification of  
the parameters \cite{Prochazka:2015deb} as 
\be
\psi_0 =  N \ , 
\ee
and 
\be
h_1 =  -\sqrt{\frac{N+k+1}{N+k}} \ , \qquad h_2 =  \sqrt{\frac{N+k}{N+k+1}} \ , \qquad h_3 = \frac{1}{\sqrt{(N+k)(N+k+1)}} \ . \label{hkn}
\ee
This calculation therefore establishes the identification of parameters proposed in \cite{Prochazka:2015deb}.

\subsection{Triality symmetry}\label{sec:triality}

Recall from the analysis of \cite{Gaberdiel:2012ku} that the triality identifications of 
the $(N,k)$ parameters are generated by the two fundamental 
transformations
\be\label{pi1}
\pi_1: N \mapsto N \ , \qquad k \mapsto -2N - k - 1 
\ee
and
\be\label{pi2}
\pi_2: N \mapsto \frac{N}{N+k} \ , \qquad k\mapsto \frac{1-N}{N+k} \ ,
\ee
see eq.~(3.10) of \cite{Gaberdiel:2012ku}. These transformations act on
the $(N,k)$ parameters as follows:
\begin{equation}
\xymatrix@C=1pc@R=2.2pc{
   &  (N,k) \ar[dl]_{\pi_1} \ar[dr]^{\pi_2}&  \\
 \textcolor{black}{(N, -1-2N-k) }    \ar[ur]   \ar[d]^{\pi_2}    & &    \ar[ul]   \ar[d]^{\pi_1}   \textcolor{black}{(\frac{N}{N+k}, \frac{1-N}{N+k})} \\ 
  \textcolor{black}{(\frac{N}{N+k}, 1-\frac{N+1}{N+k}) }    \ar[dr]    \ar[u]   & &     \ar[dl]    \ar[u]   \textcolor{black}{(-\frac{N}{N+k+1}, \frac{N-1}{N+k+1})} \\
      & \ar[ur]_{\pi_2} \ar[ul]^{\pi_1}  (-\frac{N}{N+k+1}, -\frac{k}{N+k+1})
     \nonumber
  }
\end{equation}
Under these transformations the structure constant (\ref{c433})
remains invariant, and hence the ${\cal W}_\infty$ algebra does not change. In
terms of the $\lambda$ parameters, $\pi_1$ exchanges
\be
\lambda_1\equiv \frac{N}{N+k} \quad \longleftrightarrow \quad
\lambda_2\equiv  -\frac{N}{N+k+1} \ ,
\ee
(while leaving $\lambda_3\equiv N$ invariant), 
and $\pi_2$ exchanges $\lambda_2\leftrightarrow\lambda_3$ (while leaving
$\lambda_1$ invariant).

In this section we want to understand the
incarnation of this symmetry in the language of the affine Yangian. 
Under the transformation $\pi_1$, $N$ (and hence $\psi_0$) does not change, and 
\be\label{pi1p}
\pi_1: \quad (N+k+1) \mapsto - (N+k) \ , \qquad (N+k) \mapsto - (N+k+1)  \ . 
\ee
Writing both signs as $e^{\pm \pi i}$ --- if we think of the transformation (\ref{pi1})  as arising from some analytic continuation
of $k$, both signs have the same form --- then under the action of $\pi_1$, $h_1\mapsto -h_2$, 
$h_2 \mapsto -h_1$, while $h_3 \mapsto - h_3$. The overall sign of the $h_i$ can be absorbed by the rescaling with
$\alpha=-1$ , see eq.~(\ref{halpha}), and this does not modify $\psi_0=N$, see eq.~(\ref{psialpha}).  Thus we conclude that 
the transformation $\pi_1$ acts on the $h_i$ parameters of the affine Yangian as 
\be
\pi_1 : \qquad h_1 \longleftrightarrow h_2 \ ,
\ee
while leaving $h_3$ invariant, i.e.\ as the permutation $(12)$.

The analysis for the case of $\pi_2$ is analogous. Now  $N$ is transformed to $N / (N+k)$, and hence we need to 
rescale the resulting expressions with $\alpha = (N+k)^{-1/2}$, so as to bring $\psi_0$ back to its original form. 
If we apply this $\alpha$ transformation also to the $h_i$, then one finds that $\pi_2$ corresponds to the 
transformation
\be
\pi_2:  \qquad h_2 \longleftrightarrow h_3 \ , 
\ee
while leaving $h_1$ invariant.
(For example, one finds $\pi_2(h_1) =   \sqrt{N+k+1}$, and then rescaling by 
$\alpha= (N+k)^{-1/2}$  indeed gives back $h_1$.)

The two triality transformations therefore correspond to the permutations (12) and (23) acting on the $h_i$ 
parameters, while keeping $\psi_0$ invariant; they therefore generate the full permutation group (acting on the $h_i$).
The triality symmetry acts trivially on the algebra, but exchanges the representations via the natural (geometric) 
permutation action of the $h_i$. This explains, from first principles, the observations made in \cite{Prochazka:2015deb}.

\subsection{The universal enveloping algebra and representations}\label{sec:reps}

Finally we should be a bit more precise in how the the affine Yangian of $\mathfrak{gl}_{1}$ and ${\cal W}_{1+\infty}[\lambda]$
are related to one another. Recall that the former is an associative algebra, while the latter is a commutator algebra. 
So far, we have confirmed that at least the first few generators of ${\cal W}_{1+\infty}[\lambda]$ can be expressed in terms 
of generators of the affine Yangian so that the commutation relations of ${\cal W}_{1+\infty}[\lambda]$ follow from the 
defining relations of the affine Yangian. 

We have not found a general formula for an arbitrary ${\cal W}_{1+\infty}[\lambda]$ generator in terms of 
affine Yangian generators, except for the two free field constructions that will be described in Section~\ref{sec:freefield}. 
However, we have found identifications (for general $\lambda$) for the first few ${\cal W}_{1+\infty}[\lambda]$  generators, and in 
each case the identification was of the triangular form 
\begin{align}\label{yangmap}
e_s & = \pm V^{s+1}_{-1} + \hbox{correction terms with fields of lower spin} \ , \\
f_s & = \pm V^{s+1}_{1} + \hbox{correction terms with fields of lower spin} \ .
\end{align}
In particular, this is the case for $f_0=-J_1$, $f_1=-L_1$ (as well as $e_0=J_{-1}$ and $e_1 = L_{-1}$), see
eqs.~(\ref{u1gen})  and (\ref{Virgen}). Similarly, $W^3_{-1}= - e_2 + \hbox{lower spin terms}$, see eq.~(\ref{W31}),
and $W^4_{-1} =  e_3 +\hbox{lower spin terms}$, see eq.~(\ref{W41}), and similarly for $W^3_{1}$ and
$W^4_{1}$. (The correction terms that lead to the actual local modes $V^3_{\pm 1}$ and $V^4_{\pm 1}$ are 
also of lower spin.) Thus this dictionary suggests that we can express recursively not only all ${\cal W}_{1+\infty}[\lambda]$
generators in terms of affine Yangian generators, but also conversely all $e_s$ and $f_s$ generators (and hence also all
$\psi_s$ generators) in terms of ${\cal W}_{1+\infty}[\lambda]$ generators. Since the affine Yangian algebra is the 
associative algebra generated by the modes $e_s$, $f_s$ and $\psi_s$, it then follows that the affine Yangian is isomorphic to the 
universal enveloping algebra of ${\cal W}_{1+\infty}[\lambda]$. 
\medskip

In particular, it therefore follows that the affine Yangian of $\mathfrak{gl}_1$ and ${\cal W}_{1+\infty}[\lambda]$ share the 
same representation theory. 
At least generically, i.e.\ 
as long as the vacuum representation does not possess any non-trivial null-vectors,
the representations of the vertex operator algebra associated to 
${\cal W}_{1+\infty}[\lambda]$ are  in one-to-one correspondence with those 
of the universal enveloping algebra. 
Indeed, the vertex operator algebra associates a mode to every state of the vacuum representation, and these can always be 
described in terms of normal-ordered products of monomials of the generating modes; conversely, every element of the universal 
enveloping algebra can (at least  formally) be written as a sum of modes of the vertex operator algebra. 
We therefore conclude that the representations of the vertex operator algebra associated to ${\cal W}_{1+\infty}[\lambda]$ are in 
one-to-one correspondence with those of the affine Yangian of $\mathfrak{gl}_1$. 

As explained in Section~\ref{sec:AY}, the set of plane partitions with given 
asymptotics $(\Xi^{(x)}, \Xi^{(y)} , \Xi^{(z)} )$ furnishes a 
natural representation of the affine Yangian of $\mathfrak{gl}_1$.  
A plane partition is an eigenstate of $\psi(z)$, with eigenvalue given by 
(\ref{psizdef}), and $e(z)/f(z)$ acting on it by (legally) 
adding/removing boxes, see  eqs.~(\ref{ezdef})/(\ref{fzdef}).  
Therefore the set of plane partitions with given asymptotics 
$(\Xi^{(x)} , \Xi^{(y)} , \Xi^{(z)} )$ also defines a representation of the 
$\mathcal{W}_{1+\infty}$ algebra.

Depending on how many of the three asymptotics are non-trivial, there
are four different types of plane partition representations, see Figure~\ref{fig1}. 
If at most two boundary conditions are non-trivial, and applying the triality
symmetry if necessary, we may 
assume that $\Xi^{(z)}=0$, i.e.\ that the boundary condition is described by 
$(\Xi^{(x)}, \Xi^{(y)}, 0)$. Then
the representation can be 
identified with a representation of the coset 
\be
\frac{\mathfrak{su}(N)_k \oplus \mathfrak{su}(N)_{1}}{\mathfrak{su}(N)_{k+1}}\ ,
\ee
where 
$\Xi^{(x)}$ and $\Xi^{(y)}$ are representations of the $\mathfrak{su}(N)_k$ and 
$\mathfrak{su}(N)_{k+1}$ algebra, respectively.\footnote{Here
we assume that $N$ and $k$ are sufficiently large so that the truncations that appear at finite $N$ and $k$ can be ignored.}
In the context of the dual higher spin theory, these three types correspond to  the vacuum, the perturbative states in Vasiliev theory, and the 
non-perturbative states, respectively \cite{Gaberdiel:2012ku}. 
The last type, i.e.\ the one where all three asymptotics are non-trivial, does 
not seem to have an interpretation in terms of the coset theory, i.e.\ it 
does not arise as the large $(N,k)$ limit of a coset representation. 
As far as we are aware, it defines a new type 
of representation of $\mathcal{W}_{1+\infty}$ that has not been 
constructed via any other method.

The generating function of a plane partition representation $(\Xi^{(x)} ,\Xi^{(y)} ,\Xi^{(z)} )$ counts, at level $n$, the number of ways to stack 
$n$ boxes (the yellow ones in Figure~\ref{fig1}) on top of the ground state, given by the minimal configuration obeying the boundary condition 
$(\Xi^{(x)} ,\Xi^{(y)} ,\Xi^{(z)} )$ (the blue configuration of Figure~\ref{fig1}). 
For the coset type representations (i.e.\ for  $\Xi^{(z)} =0$), the plane partition generating function is identical to the character computed via the 
Ka\u c-Weyl formula \cite{FFJMM2}. One advantage of the plane partition viewpoint 
in describing representations of $\mathcal{W}_{\infty}$, even 
those of the coset type,  is that it is much easier to compute the character via the combinatorics of box stacking than using the 
Ka\u c-Weyl character formula. For example, this idea was used to identify the twisted sector representations 
of the symmetric orbifold in \cite{Datta:2016cmw}.

The plane partition representations are quasi-finite, i.e.\ there are only finitely many states at each level.  
We note that the affine Yangian (as well as ${\cal W}_{1+\infty}[\lambda]$) also possess larger representations; 
for example, for $\lambda=N$ there are also representations that are labelled by $N$ independent Young diagrams. 
(This follows from the fact that the algebra SH$^c$, which is 
isomorphic to the affine Yangian of $\mathfrak{gl}_1$, see \cite{Zhu:2015nha} and Section~\ref{sec:SHc} below, has such representations, see e.g.\ \cite{Kanno:2013aha}.)

\section{Free field realizations}\label{sec:freefield}

For $\lambda=0$ and $\lambda=1$, the ${\cal W}_\infty[\lambda]$ algebra has a free field construction
in terms of free fermions and free bosons, respectively. Thus we should expect that, for these values of $\lambda$,
we can find closed-form expressions for the affine Yangian generators in terms of the corresponding free fields.
For the case of $\lambda=0$ this description was already found in \cite{Prochazka:2015deb} (and we shall only
briefly review it below), while the construction for $\lambda=1$ appears to be new. The fact that at least for
these special values of $\lambda$ we can establish a closed-form dictionary between the affine Yangian and
the ${\cal W}_\infty$ generators gives strong support to the idea that the partial dictionary we established in Section~\ref{sec:comp}
can be generalized to all spin fields. 

Before we describe the details for the two cases, we would like to make one general comment.
The case $\lambda=0$ correspond to taking $k\rightarrow \infty$ at fixed $N$, while $\lambda=1$ is described
by taking $N\rightarrow \infty$ at fixed $k$. In either case, $\sigma_2$ and $\sigma_3$,
defined by eq.~(\ref{sigma23}), become
\begin{equation}
\begin{aligned}
\s_2    
& = -1 - \frac{1}{(N+k)(N+k+1)}  \cong -1 \ , \\ 
\sigma_3 
& = - \frac{1}{\sqrt{(N+k)(N+k+1)}}  \cong 0  \ . 
\end{aligned}
\end{equation}
In particular, the non-local correction terms of Section~\ref{sec:comp} and Appendix~\ref{app:Spin4}
are absent since they are proportional to $\sigma_3$, see eqs.~(\ref{defV}) and (\ref{V4def}). 
Furthermore, at least formally, the anti-commutator terms in the definition of the affine Yangian in eqs.~(\ref{Y1}), (\ref{Y2}), 
(\ref{Y4}) and (\ref{Y5}) drop out. However, there is one important subtlety: in the relations 
eqs.~(\ref{Y4}) and (\ref{Y5}), $\sigma_3$ is multiplied by $\psi_j$, and for $j=0$, $\psi_0 = N$
is taken to infinity for $\lambda=1$. Hence while for $\lambda=0$ all of these anti-commutator terms are indeed
absent, for $\lambda=1$, the expression $\sigma_3 \psi_0$ becomes 
\be
\sigma_3 \psi_0 = -\frac{N}{\sqrt{(N+k)(N+k+1)}} \cong -1  \ ,
\ee
and hence leads to a correction term for the special cases
\begin{align}
[\psi_3,e_k] &= 3 [\psi_2,e_{k+1}] - 3 [\psi_1, e_{k+2}] + [\psi_0, e_{k+3}] +[\psi_1,e_k] - [\psi_0,e_{k+1}] - 2  e_k \nonumber \\
& =  6 e_{k+1}  - 2 e_k  \ ,\label{inim1}
\end{align}
and
\begin{align}
[\psi_3,f_k] &= 3 [\psi_2,e_{k+1}] - 3 [\psi_1, e_{k+2}] + [\psi_0, e_{k+3}] +[\psi_1,e_k] - [\psi_0,e_{k+1}] + 2  f_k \nonumber \\
& =  - 6 f_{k+1}  + 2 f_k  \ . \label{inim2}
\end{align}
This will be important in our construction below. 

\subsection{The free fermion construction}

The construction for $\lambda=0$ was already given in \cite{Prochazka:2015deb}, and hence we shall be brief. 
We start with $N$ free complex fermions $\psi^i$ and $\bar\psi^i$ with $i=1,\ldots,N$. The bilinear ${\rm U}(N)$
singlets (where we take the $\psi^i$ to transform in the fundamental representation of ${\rm U}(N)$, and 
the $\bar\psi^i$ in the anti-fundamental) generate the linear ${\cal W}_{1+\infty}$ algebra
\cite{Bergshoeff:1989ns,Bergshoeff:1990yd,Depireux:1990df}, see also \cite{Gaberdiel:2013jpa}. Thus 
the Yangian generators should also be expressed in terms of such bilinears, and one finds that 
\begin{equation}
\begin{aligned}
\psi_r & =  \sum_{m\in\mathbb{Z}+\frac{1}{2}} \sum_{i=1}^{N} \Bigl( (-m-\tfrac{1}{2})^r - (-m+\tfrac{1}{2})^r \Bigr)\, :\bar{\psi}^i_{-m}\psi^i_m : \ , \\
f_s  & =  \sum_{m\in\mathbb{Z}+\frac{1}{2}} \sum_{i=1}^{N} \bigl(-m+\tfrac{1}{2})^s\, :\bar{\psi}^i_{-m+1} \psi^i_m : \ , \\
e_s & = - \sum_{m\in\mathbb{Z}+\frac{1}{2}} \sum_{i=1}^{N}  \bigl(-m-\tfrac{1}{2})^s\, :\bar{\psi}^i_{-m-1} \psi^i_m : 
\end{aligned}
\end{equation}
satisfy all the relations of the affine Yangian at $\sigma_3=0$, $\sigma_2=-1$. 
Note that the definition of $\psi_0$  formally vanishes, but that in order to 
reproduce the correct central charge $\tilde{c}=N-1$ in (\ref{c1})
we should set  $\psi_0 = N$.  (Since $\psi_0$ is central and since all anti-commutators
drop out, we are free to set $\psi_0$ to any value we chose without modifying the defining relations of the affine Yangian.) 

One can also check that the above generators give rise to the correct $W$-generators using our general identification between the 
${\cal W}_\infty$ generators and the affine Yangian generators. For example, we have from eq.~(\ref{W31m})  
\be
W^3_{1} = f_2 + \tfrac{1}{2} \sigma_3 \psi_0\, f_1 \cong f_2 
\ee
in the $\lambda\rightarrow 0$ limit, leading to 
\be
W^3_1 = \sum_{m\in\mathbb{Z}+\frac{1}{2}} \sum_{i=1}^{N}  \, \bigl(m-\tfrac{1}{2})^2\, :\bar{\psi}^i_{-m+1} \psi^i_m : \ , 
\ee
which reproduces the usual free field answer, see e.g.\ eq.~(2.8) of \cite{Gaberdiel:2013jpa}. (Recall that at 
$\sigma_3=0$ there is no difference between the $V^3_n$ 
and the $W^3_n$ generators, see eq.~(\ref{defV}).)

\subsection{The free boson construction}

The situation for the free boson case is slightly more subtle. We start again with $k$ free complex boson fields,
whose modes satisfy the commutation relations
\be
{}[\alpha^i_m,\bar\alpha^j_n] = m \, \delta^{ij}\, \delta_{m,-n} \ , 
\ee
while all other commutators vanish. The ${\rm U}(k)$ singlets (where the 
$\alpha^i_m$/$\bar\alpha^i_m$ transform in the 
fundamental/anti-fundamental representation of ${\rm U}(k)$)
generate now only a linear ${\cal W}_\infty$ algebra, and 
there is no spin-one generator. As a consequence,
we should only expect to be able to express the affine Yangian generators $e_r$, $f_r$ with $r\geq 1$ and $\psi_s$ with $s\geq 2$
in terms of these free fields; note that the algebra generated by this subset of fields is a well-defined subalgebra of the affine Yangian
algebra.\footnote{It is also worth pointing out that this subset of Yangian generators will only correspond to the wedge
subalgebra of ${\cal W}_\infty$ --- all the modes below will be from within the wedge.}
 For these generators we make the ansatz 
\begin{equation}
\begin{aligned}
f_r & = - \sum_{m\in\mathbb{Z}} \sum_{j=1}^{k} \bigl(-m+1\bigr)^{r-1}\, : \alpha^j_m  \bar\alpha^j_{1-m} : \\
e_r & = \sum_{m\in\mathbb{Z}}  \sum_{j=1}^{k} (- m)^{r-1}\, : \alpha^j_m  \bar\alpha^j_{-1-m} :  \\
\psi_r & =   \sum_{m\in\mathbb{Z}}  \sum_{j=1}^{k} \Bigl( (  m + 1)(-m)^{r-2} + \bigl(-m+1 \bigr)^{r-1} \Bigr)\, 
: \alpha^j_{m} \bar\alpha^j_{-m}: \ ,
\end{aligned}
\end{equation}
where the $\psi_r$ modes are determined by the condition  
$[e_r, f_s] = \psi_{r+s}$, see eq.~(\ref{Y3}).
Note that with respect to the hermitian structure defined by $(\alpha_m^{j})^{\dagger} =  \bar\alpha_{-m}^j$, we have 
$e_r^\dagger = -f_r$. These modes then satisfy all relations of the affine Yangian algebra. In particular, the commutators 
of two $e_r$ generators are 
\be
{}[f_r,f_s] = \sum_{m\in\mathbb{Z}}  \sum_{j=1}^{k} \Bigl(\, \bigl( - m +1 \bigr)^{r}  \bigl( -m +2 \bigr)^{s-1}   -\bigl( - m + 2 \bigr)^{r-1} \bigl( - m +1\bigr)^{s} \Bigr)
\, 
: \alpha^j_m  \bar\alpha^j_{2-m} : 
\ee
and similarly 
\be
{}[e_r,e_s] = \sum_{m\in\mathbb{Z}}  \sum_{j=1}^{k}  \Bigl( ( - m)^{r-1}\,  \bigl(-m-1\bigr)^{s} \, - \bigl(-m-1\bigr)^{r}\,  ( - m )^{s-1}  \Bigr) \, : \alpha^j_m  \bar\alpha^j_{-2-m} :  \ .
\ee
It is then straightforward to check that they satisfy the relations 
(\ref{Y0}) -- (\ref{Y4}) with $\sigma_2=-1$ and $\sigma_3=0$, as well as (\ref{Serre}). 
One also confirms directly that they indeed give rise to the correct initial conditions 
\begin{eqnarray}
&[\psi_2,f_{k}] = -2 f_k \ , \qquad
& [\psi_3,f_{k}] = - 6 f_{k+1}+2 f_k \ , \\
&[\psi_2,e_{k}] = 2 e_k \ , \qquad
& [\psi_3,e_{k}] = 6 e_{k+1} - 2 e_k \ .
\end{eqnarray}
Here we have used that since the $\psi_s$  are only defined for $s\geq 2$, only the last equation from eq.~(\ref{ini1}) and (\ref{ini2}) makes sense, 
and because of  (\ref{inim1}) and (\ref{inim2}), the commutator with $\psi_3$ has indeed the required form. 

Again, we can also check that the above generators give rise to the correct
$W$-generators using our general identification between the ${\cal W}_\infty$ generators and the affine
Yangian generators. For example, we have from eq.~(\ref{W31m}) 
\be
W^3_{1} = f_2 + \tfrac{1}{2} \sigma_3 \psi_0\, f_1 \cong f_2 - \tfrac{1}{2} f_1  \ ,
\ee
where we have taken the $\lambda\rightarrow 1$ limit. This leads to 
\be
W^3_{1} = \sum_{m\in\mathbb{Z}}\,   \sum_{j=1}^{k}  \, (m-\tfrac{1}{2})\, :\alpha^j_m \bar\alpha^j_{1-m} : \ , 
\ee
which reproduces (up to an overall normalization factor) the usual free field answer, 
see for example eq.~(2.5) of \cite{Gaberdiel:2015wpo}.
(Again, at $\sigma_3=0$ there is no difference between the 
$V^3_n$ and $W^3_n$ generators, see eq.~(\ref{defV}).)

\section{The relation to SH$^c$}\label{sec:SHc}

The affine Yangian is  believed to be isomorphic \cite{Zhu:2015nha,Prochazka:2015deb} to the spherical degenerate double affine Hecke algebra, 
the so-called SH$^c$ algebra of \cite{SV}, although the detailed dictionary has, to our knowledge, not been written down before. 
In this section we exhibit this isomorphism in detail; we also 
explain how this fits together with the equivalence to the universal enveloping algebra of ${\cal W}_{1+\infty}[\lambda]$.

\subsection{The SH$^c$ algebra and its isomorphism to the affine Yangian 
of $\mathfrak{gl}_1$}

The definition of the SH$^c$ algebra is spelled out in Def.~1.31 of \cite{SV}; here we follow the description
of the algebra in terms of generating functions that was worked out in \cite{Bourgine:2014tpa,Bourgine:2015szm}.
The definition of SH$^c$ is not manifestly triality invariant;
in order to rectify this, it is convenient to choose an arbitrary parameter $h_1$, and define $h_2$ and $h_3$, using
the parameters $\kappa$ and $\xi$ that appear in \cite{SV} via
\be\label{kappadef}
\kappa = - \frac{h_2}{h_1} \ , \qquad \xi = 1 - \kappa = - \frac{h_3}{h_1} \ . 
\ee
Then we introduce the generating functions\footnote{The following definition is a slight correction of the 
conventions used in \cite{Bourgine:2015szm}.}
\begin{equation}\label{Dgen}
{\cal D}_{\pm 1}(z) \equiv \sum_{j=0}^{\infty} \, \frac{(h_1)^j\, D_{\pm 1, j}}{z^{j+1}} \qquad
\textrm{and} \qquad 
{\cal D}_0(z)  \equiv \sum_{j=0}^{\infty}  \frac{(h_1)^j\, D_{0,j+1}}{z^{j+1}} \ . 
\end{equation}
The algebraic relations of SH$^c$ are then 
\begin{eqnarray}
{}[{\cal D}_0(z), {\cal D}_0(w)] & = & 0 \\
{}[{\cal D}_0(z), {\cal D}_{\pm 1} (w)] & = & \mp \frac{{\cal D}_{\pm 1}(z)- {\cal D}_{\pm 1}(w)}{z-w}  \label{D0D1} \\
{}[{\cal D}_{+1}(z),{\cal D}_{-1}(w)]& = & - \frac{1}{h_3} \frac{{\cal E}(z)-{\cal E}(w)}{z-w} \ ,  \label{DpDm}
\end{eqnarray}
where ${\cal E}(z)$ is the generating function of the modes $E_j$
\begin{equation}\label{Egen}
{\cal E}(z)=1 - h_3 \sum_{j=0} \frac{(h_1)^j\, E_j}{z^{j+1}} \ ,
\end{equation}
which are related in turn to the $D_{0,j}$ modes by eq.~(1.73) of \cite{SV}. To express this relation more
conveniently, we introduce ${\cal X}(z)$ via \cite{Bourgine:2015szm}
\begin{equation}
	{\cal X}(z)\equiv \int^z dz' \, {\cal D}_{0}(z')= 
	D_{0,1}\, \log \bigl({\frac{z}{h_1}}\bigr) - \sum^{N}_{j=1} \frac{1}{j}\, \frac{(h_1)^j\, D_{0,j+1}}{z^{j}}  \ , 
\end{equation}
and then define ${\cal Y}(z)$ as 
\begin{equation}
{\cal Y}(z)\equiv e^{c(z)}\frac{e^{{\cal X}(z-h_1)}e^{{\cal X}(z-h_2)}}{e^{{\cal X}(z)}e^{{\cal X}(z+h_3)}} \ , 
\end{equation}
where $c(z)$ is the generating function of the central charges
\be
c(z) = \sum_{j=0}^{\infty} \frac{(h_1)^j\, c_j}{z^{j+1}} \ .
\ee
Then we have the simple relation 
\begin{equation}\label{Edef}
{\cal E}(z)= \frac{{\cal Y}(z-h_3)}{{\cal Y}(z)} \ .
\end{equation}
\smallskip

\noindent It was shown in \cite{Bourgine:2015szm} that, on the $N$-tuple Young diagram representations (that
form a faithful representation of SH$^c$, see Corollary 8.7 of \cite{SV}) the 
SH$^c$ relations imply, see eq.~(2.30) of that paper, 
\begin{equation}\label{YDrel}
{\cal Y}(z)\, {\cal D}_{+1}(w)\, \frac{1}{{\cal Y}(z)}=g(z-w+h_3) \, {\cal D}_{+1}(w) 
+\frac{h_1 h_2}{h_3} \left[\frac{{\cal D}_{+1}(z+h_3)}{z-w+h_3}-\frac{{\cal D}_{+1}(z)}{z-w}\right] \ , 
\end{equation}
where $g(z)$ is defined via 
\begin{equation}
g(z)\equiv\frac{(z+h_1)(z+h_2)}{z(z-h_3)} \ .
\end{equation}
Note that the second term in (\ref{YDrel}) only corrects for the poles that are explicitly introduced by the 
function $g(z)$; in particular, multiplying the whole equation by $(z-w+h_3) (z-w)$ we obtain 
\begin{eqnarray}
(z-w+h_3) (z-w) \, {\cal Y}(z)\, {\cal D}_{+1}(w)\, \frac{1}{{\cal Y}(z)} & = & 
(z-w-h_2) (z-w-h_1) \, {\cal D}_{+1}(w) \nonumber \\
& &  + \frac{h_1 h_2}{h_3} \, (z-w)\, {\cal D}_{+1}(z+h_3)  \\
& & -  \frac{h_1 h_2}{h_3} \, (z-w+h_3) \, {\cal D}_{+1}(z) \  .  \nonumber 
\end{eqnarray}
The left-hand-side vanishes for $z=w$, while the right-hand-side then becomes 
\be
(-h_2) (-h_1)  {\cal D}_{+1}(w) - \frac{h_1 h_2}{h_3} \, h_3 \, \left. {\cal D}_{+1}(z) \right|_{z=w} = 0 \ , 
\ee
because of the second term. The other correction term similarly guarantees that the equations holds for $z=w-h_3$. 
Using (\ref{Edef}), eq.~(\ref{YDrel}) then leads to the identity 
\begin{equation}
\begin{aligned}
{\cal E}(z)\, {\cal D}_{+1}(w)\, \frac{1}{{\cal E}(z)} = 
& \,\varphi (z-w) \, {\cal D}_{+1}(w)\\
&  +\frac{2h_1 h_2 h_3}{(h_1-h_2)(h_2-h_3)(h_3-h_1)} 
\Bigl[(h_2-h_3)\frac{{\cal D}_{+1}(z-h_1)}{z-w-h_1} \\
&  \qquad +(h_3-h_1)\frac{{\cal D}_{+1}(z-h_2)}{z-w-h_2} +(h_1-h_2)
\frac{{\cal D}_{+1}(z-h_3)}{z-w-h_3} \Bigr] \ . 
\end{aligned}
\end{equation}
Again, the terms of the second and third line only correct for the poles that are explicitly introduced by
$\varphi(z-w)$ at $z= w+h_i$. Since these do not contribute to the terms that have {\em both} negative Fourier coefficients
in $z$ and $w$, which are the only terms we are interested in, 
given the definition of the generating functions
(\ref{Dgen}) and (\ref{Egen}), we can write this identity as 
\be
{\cal E}(z)\, {\cal D}_{+1}(w)\, \sim \varphi (z-w) \, {\cal D}_{+1}(w) \, {\cal E}(z) \ ,
\ee
where $\varphi(z-w)$ is the same function as defined in (\ref{varphidef}). 
Similarly, we find for ${\cal D}_{-1}$ the relation
\be
{\cal E}(z)\, {\cal D}_{-1}(w)\, \sim \varphi (w-z) \, {\cal D}_{-1}(w) \, {\cal E}(z) \ . 
\ee
These identities now look very similar to the relations that appear in the definition of the affine Yangian, see
in particular eqs.~(\ref{psiegen}) and (\ref{psifgen}). (The relation eq.~(\ref{DpDm}) is also essentially
the same as eq.~(\ref{efgen}). On the other hand, the relations eqs.~(\ref{eegen}) and (\ref{ffgen}) give
rise to commutation relations for the higher modes that are implicitly defined in eq.~(2.4) of  \cite{Bourgine:2015szm}.)
Thus the isomorphism of the two algebras simply amounts to the identification
\be\label{shcyang}
e(z) = \frac{1}{\sqrt{h_1 h_2}}\, {\cal D}_{+1}(z) \ , \qquad f(z) = \frac{1}{\sqrt{h_1 h_2}}\, {\cal D}_{-1}(z) \ , \qquad 
\psi(z) = {\cal E}(z) \ . 
\ee
In terms of modes, this means that we have the identification
\begin{equation}\label{modeiden}
e_j = \frac{(h_1)^j}{\sqrt{h_1 h_2}}\, D_{1,j}  \ , \qquad 
f_j = \frac{(h_1)^j}{\sqrt{h_1 h_2}}\, D_{-1,j} \ , \qquad
\psi_j = - \frac{ (h_1)^j}{h_1 h_2}\, E_{j} \ . 
\end{equation}

\subsection{The relation to ${\cal W}_{1+\infty}[\lambda]$}

It is argued in \cite{SV}, 
see Remark 8.28, that the SH$^c$ algebra is isomorphic to the universal enveloping algebra of ${\cal W}_{1+N}$ realized via a 
Drinfeld-Sokolov construction with level 
\be\label{kDS}
k_{\rm DS}= \kappa - N \ , 
\ee
where $\kappa$ is the parameter that appears in the definition of the SH$^c$. 
The relation of the DS level $k_{\rm DS}$ and the usual coset level  $k$ is, see e.g., \cite[eq.~(7.52)]{Bouwknegt:1992wg} 
\be\label{kDSrel}
\frac{1}{k_{\rm DS} + N } 
= \frac{1}{k+N} + 1 \ ,
\ee
from which we conclude that the $\kappa$ parameter in SH$^c$ equals, in terms of the $(N,k)$ parametrization
\be\label{kappadef1}
\kappa = \frac{k+N}{k+N+1}  = - \frac{h_2}{h_1}\ ,
\ee
where we have used (\ref{hkn}) in the last step. This then precisely agrees with (\ref{kappadef}). 
In terms of the parameters of the affine Yangian, the translation of the parameters is 
\be\label{dic1}
\kappa + \kappa^{-1} - 2 = - \frac{(h_3)^2}{ h_1 h_2} \ . 
\ee
Note that the expression on the RHS is invariant under the scaling symmetry
(\ref{halpha}).

The triality symmetry of the affine Yangian, which we studied in Section~\ref{sec:triality},
has also an incarnation for the 
case of SH$^c$; this was already studied in \cite{Fukuda:2015ura}.

%%%%%%%%%%%%%%%%
\subsection{$N$-tuple Young diagram representation of $\text{SH}^c$}
%%%%%%%%%%%%%%%%%

The SH$^c$ algebra acts naturally on a vector space whose basis 
vectors are labelled by an $N$-tuple of Young diagrams 
$\vec{\lambda} \equiv (\lambda_1,\ldots,\lambda_N)$. Here each $\lambda_i$ is a Young diagram, and the 
representation is characterized by a vector of complex numbers $\vec{a} = (a_1,\ldots,a_N)$. More specifically,
each $N$-tuple Young diagram $\vec{\lambda}$ is an eigenvector of the operator ${\cal E}(z)$
\be
{\cal E}(z) \, |\vec{\lambda}\rangle_{\vec{a}} = {\cal E}_{\vec{\lambda}}(z) \, |\vec{\lambda}\rangle_{\vec{a}} \ , 
\ee
where the eigenvalue equals
\be
{\cal E}_{\vec{\lambda}}(z) = \prod_{ {\tiny \yng(1)} \in {\rm Add}(\vec{\lambda})}
\Bigl( 1 - \frac{h_3}{z-\phi({\tiny \yng(1)})} \Bigr) \, 
 \prod_{{\tiny \yng(1)}\in {\rm Rem}(\vec{\lambda})}  \Bigl( 1 + \frac{h_3}{z-\phi({\tiny \yng(1)})} \Bigr) \ .
\ee
Here ${\rm Add}(\vec{\lambda})$ is the set of boxes that can be added to the $N$ Young diagrams
(so that the resulting configuration still describes an $N$-tuple of Young diagrams), while 
${\rm Rem}(\vec{\lambda})$ is the set of boxes that can be consistently removed.
Furthermore, the function 
$\phi({\tiny \yng(1)})$ is defined via
\be\label{phidef}
\phi({\tiny \yng(1)}) = a_{i({\tiny \yng(1)})} + h_1 x({\tiny \yng(1)}) + h_2 y({\tiny \yng(1)}) \ , 
\ee
where $i({\tiny \yng(1)})$ denotes which of the $N$ Young diagram the box is associated to, while 
$x({\tiny \yng(1)})$ and $y({\tiny \yng(1)})$ are its $(x,y)$ coordinate --- here 
the Young diagrams lie in the $xy$-plane, with the first box having coordinates
$(x,y)=(0,0)$, and the different Young diagrams are lined up along the $z$-direction.
(The alert reader
will notice that this is a generalization of (\ref{hbox}), to which it reduces if $a_i = h_3 (i-1)$.) 
With these conventions,
the action of ${\cal D}_{\pm 1}(z)$ on these states is defined as
\be\label{5.25}
{\cal D}_{+ 1}(z) \, |\vec{\lambda}\rangle_{\vec{a}} = 
\sum_{ {\tiny \yng(1)} \in {\rm Add}(\vec{\lambda})} 
\frac{\bigl[ -  \frac{1}{h_3} {\rm Res}_{w=\phi({\tiny \yng(1)})} {\cal E}_{\vec{\lambda}}(w) \bigr]^{\frac{1}{2}}}{ z - \phi({\tiny \yng(1)}) } \, 
|\vec{\lambda} + {\tiny \yng(1)} \rangle_{\vec{a}}  \ , 
\ee
\be\label{5.26}
{\cal D}_{- 1}(z) \, |\vec{\lambda}\rangle_{\vec{a}} = 
\sum_{ {\tiny \yng(1)} \in {\rm Rem}(\vec{\lambda})} 
\frac{ \bigl[ -  \frac{1}{h_3} {\rm Res}_{w=\phi({\tiny \yng(1)})} {\cal E}_{\vec{\lambda}}(w) \bigr]^{\frac{1}{2}}}{ z - \phi({\tiny \yng(1)})} \, 
|\vec{\lambda} - {\tiny \yng(1)} \rangle_{\vec{a}}  \ , 
\ee
where ${\rm Res}$ is the residue, while ${\cal D}_0(z)$ acts diagonally as 
\be
{\cal D}_{0}(z) \, |\vec{\lambda}\rangle_{\vec{a}} =  \sum_{ {\tiny \yng(1)} \in \vec{\lambda}}\, \frac{1}{z - \phi( {\tiny \yng(1)})} \, 
|\vec{\lambda}\rangle_{\vec{a}} \ . 
\ee
Here the $a_i$ are treated as formal (independent) variables, and it follows by the same arguments as in 
\cite{SV} (see also \cite{Bourgine:2015szm}) that this defines a representation of SH$^c$. 

For generic $a_i$, there are $N$ states at level one. To obtain more 
special representations, in particular those with many null states, we 
need to choose special $a_i$. The most extreme example of this is the
vacuum representation of ${\cal W}_{1+\infty}$, for which we choose 
(cf.\ the comment below eq.~(\ref{phidef}))
\be
a_i = a + h_3 (i-1) \ .
\ee
It is not difficult to see that the numerator factor in (\ref{5.25}) guarantees that one cannot add a box to the 
$(i+1)^{\rm th}$ Young diagram at position $(x,y)$, if one could also add a box at position $(x,y)$ to the $i^{\rm th}$.
Thus, starting from the configuration of $N$ empty Young diagrams, one can only add a box to the $(i+1)^{\rm th}$ Young diagram
if the corresponding position is already occupied in the $i^{\rm th}$. The resulting configurations are therefore
precisely the plane partitions, whose counting function agrees with the MacMahon function (and 
thus the vacuum character of ${\cal W}_{1+\infty}$). 

Similarly, the representations whose non-trivial asymptotics are described by the single Young diagram $\Xi^{(x)} $ are obtained upon
choosing
\be\label{cond1}
a_{i+1} = a_i + h_3 - h_1 (\Xi^{(x)} _{i} - \Xi^{(x)} _{i+1}) \ , 
\ee
where $\Xi^{(x)} _j$ is the number of boxes in the $j^{\rm th}$ row of $\Xi^{(x)} $. 
Indeed, then the numerator factor in (\ref{5.25}) 
guarantees that one cannot add a box to the $(i+1)^{\rm th}$ Young 
diagram at position $(x,y)$, if one could also add a box 
at position $(x-n,y)$ to the $i^{\rm th}$ Young diagram, where 
$n=\Xi^{(x)} _{i}-\Xi^{(x)} _{i+1}$. For $n>0$ it is therefore possible
to add a box to the $(i+1)^{\rm th}$ Young diagram
at $(x,y)$, even if the corresponding position has not yet been filled in the 
$i^{\rm th}$; effectively, this is equivalent to imposing a non-trivial asymptotic 
(described by $\Xi^{(x)} $) for the first
few Young diagrams, i.e., the resulting configurations are counted by 
plane partitions with boundary condition $\Xi^{(x)} $. By the same reasoning, 
for the case where there are 
non-trivial boundary conditions both
in the $x$- and the $y$-direction (described by the Young diagrams 
$\Xi^{(x)}$ and $\Xi^{(y)}$,
respectively), the relevant formula becomes 
\be\label{cond2}
a_i = a + h_3 (i-1) - h_1\, \Xi_i^{(x)} - h_2\, \Xi_i^{(y)} \ .
\ee
We should note that the action of ${\cal D}_{+1}(z)$, see eq.~(\ref{5.25}), is essentially the same as 
that for the corresponding operator $e(z)$, see eq.~(\ref{ezdef}); in particular, apart from the $h_1$- and $h_2$-terms
that effectively implement the non-trivial asymptotics (see above), (\ref{phidef}) with $a_i$ being defined by 
(\ref{cond2}) agrees with (\ref{hbox}).

\section{Discussion}\label{sec:conclusion}

In this work we have spelled out some of the relations between the ${\cal W}_{\infty}$ algebra and certain novel algebraic structures 
that have been studied in recent years such as the  affine Yangian of $\mathfrak{gl}_1$ 
and the SH$^c$ algebra. We expect 
these alternative viewpoints to shed new light on both the ${\cal W}_{\infty}$ algebra as well as the Yangians. We have already 
seen that the triality symmetry of ${\cal W}_{\infty}$, which is not obvious in any of its conventional formulations, is manifest in the 
Yangian description. The Yangian picture is also very natural in the study of the degenerate representations of 
${\cal W}_{\infty}$ in terms of plane partitions. 

We can also expect insight in the reverse direction. 
Yangian symmetries have played a role in understanding the two-dimensional integrable structures underlying planar gauge theories 
such as ${\cal N}=4$ super Yang-Mills.  These symmetries act non-locally on the dual worldsheet description, which makes 
it difficult to tease out their consequences. As noted above, 
the map \eqref{yangmap} between generic 
Yangian generators and those of ${\cal W}_{\infty}$ is non-local. 
This suggests that we might be able to use (an analogue of) this 
relation, or rather, its inverse, to search for an alternative, local worldsheet description to the Yangians that appear in 
integrable spin chains \cite{Torrielli:2010kq}. It would also be particularly interesting to connect this to the work on integrability in 
${\rm AdS}_3/{\rm CFT}_2$ \cite{OhlssonSax:2011ms, Sax:2012jv, Borsato:2015mma, Sfondrini:2014via}.   

One of the original motivations for this work was to get a better handle on the unbroken stringy 
symmetries of ${\rm AdS}_3$ backgrounds, as captured by dual symmetric 
product CFTs, which
are much bigger than ${\cal W}_{\infty}$.  It was shown in 
\cite{Gaberdiel:2015mra} that for $\lambda=0$ and $\lambda=1$ the bosonic 
${\cal W}_\infty[\lambda]$ algebra can be extended to a much larger symmetry algebra, the 
analogue of the stringy ${\cal W}$-algebra of the ${\cal N}=4$ superconformal case 
\cite{Gaberdiel:2014cha}. 
Given that the stringy ${\cal W}$-algebra contains the ${\cal W}_\infty$ algebra as a subalgebra, it 
is natural to ask how it is related to the affine Yangian (or the SH$^c$ algebra). Unfortunately, we have not 
been able to find a direct relation between these structures so far. The stringy symmetry algebra is equivalent
to the universal enveloping algebra of a complex boson theory --- this is simply the familiar fact that the symmetry algebra
of a symmetric orbifold is generated, in the large $N$ limit, by the (single-particle) generators that are in one-to-one
correspondence with all the states (including the multi-particle states) of the underlying seed theory. However, this 
identification does not  seem to give rise to any simple relation between the stringy symmetry algebra and 
the universal enveloping algebra of ${\cal W}_{1+\infty}$, i.e., the affine Yangian. On the other hand, the structure
of the stringy symmetry algebra is fairly reminiscent of a Yangian algebra \cite{Gaberdiel:2015mra}, and this could be a sign 
that yet another Yangian algebra plays an important role here.

\section*{Acknowledgements} We are grateful to Akash Goel, Vasily Pestun, Tomas Prochazka, and
Hong Zhong for discussions and correspondences. 
MRG thanks ITP, Chinese Academy of Sciences, where part of this work was done, for hospitality.
The work of CP was supported by a grant from the Swiss National Science Foundation and also partly by the US Department of Energy under contract DE-SC0010010 Task A. RG is supported partly by a J. C. Bose Fellowship and more broadly by the support for basic science from the people of India.
MRG and WL thank the first Mandelstam School for Theoretical Physics for hospitality, and CP appreciates the hospitality of ETH Zurich, KU Leuven, and the Universit\'{e} Libre de Bruxelles during the final stages of this work. This research was also (partly) supported by the NCCR SwissMAP, funded by the Swiss National Science Foundation.

\appendix

\section{The construction of the local spin-$4$ field}\label{app:Spin4}

For the wedge modes of the spin-$4$ field we make the ansatz
\begin{align}
W^4_{-3}&=-\tfrac{1}{2}[e_1,[e_1,e_3]]\\
W^4_{-2}&=-\tfrac{1}{2}([e_1,e_3]+\s_3\psi_0 [e_1,e_2]) \label{W42}\\
W^4_{-1}&=e_3+\s_3\psi_0 e_2-\tfrac{1}{5}(\s_2-\s_3\psi_1-\s_3^2\psi_0^2)e_1 \label{W41}\\
W^4_0&=\tfrac{1}{4}\psi_4+\tfrac{1}{4} \s_3\psi_0 \psi_3 -\tfrac{1}{20} (\s_2-\s_3\psi_1-\s_3^2\psi_0^2)\psi_2\\
W^4_{ 1}&=-f_3-\s_3\psi_0 f_2+\tfrac{1}{5}(\s_2-\s_3\psi_1-\s_3^2\psi_0^2)f_1\\
W^4_{ 2}&=\tfrac{1}{2}([f_1,f_3]+\s_3\psi_0 [f_1,f_2])\\
W^4_{ 3}&=-\tfrac{1}{2}[f_1,[f_1,f_3]]\ .
\end{align}
Again, they satisfy the correct commutation relations with the wedge modes of the Virasoro algebra
\be
[L_m, W^4_n]=(3m-n) W^4_{m+n}\qquad \hbox{for $m=0,\pm 1$} \ .
\ee
Some examples of their commutators with the spin-one generators are
\begin{align}
[J_{-1},W^4_3]&=3 W^3_2\\
[J_{-2},W^4_3]&=6 W^3_1-3\s_3\psi_0 L_1\\
[J_{-3},W^4_3]&=9 W^3_{0}-9\s_3\psi_0 L_0 +3 (\s_2- \sigma_3^2 \psi_0^2-\s_3 \psi_1) J_0\\
[J_{-1},W^4_2]&=3 W^3_1-\tfrac{1}{2} \s_3 \psi_0 L_1 \\
[J_{-2},W^4_2]&=6 W^3_{0}-4\s_3\psi_0 L_0 + (\s_2-\sigma_3^2 \psi_0^2 -\s_3 \psi_1)J_0\\
[J_{-3},W^4_2]&=9 W^3_{-1}-\tfrac{21}{2}\s_3\psi_0 L_{-1} +6 (\s_2- \sigma_3^2\psi_0^2-\s_3 \psi_1)J_{-1}\\
[J_{-1},W^4_1]&=3 W^3_{0}-\s_3\psi_0 L_0 +\tfrac{1}{5} (\s_2-\sigma_3^2\psi_0^2 -\s_3 \psi_1)J_0\\
[J_{-2},W^4_1]&=6 W^3_{-1}-5\s_3\psi_0 L_{-1} +\tfrac{12}{5} (\s_2- \sigma_3^2 \psi_0^2-\s_3 \psi_1)J_{-1}\\
[J_{-3},W^4_1] &=9 W^3_{-2}-12\s_3\psi_0 L_{-2} +\tfrac{48}{5} (\s_2- \sigma_3^2\psi_0^2-\s_3 \psi_1)J_{-2}{-6\s_3 (J_{-1}J_{-1}-J_0 J_{-2})} \nonumber \\
[J_{-1},W^4_0]&=3 W^3_{-1}-\tfrac{3}{2}\s_3\psi_0 L_{-1} +\tfrac{3}{5} (\s_2- \sigma_3^2\psi_0^2-\s_3\psi_1)J_{-1}\\
[J_{-2},W^4_0] &=6 W^3_{-2}-6 \s_3\psi_0 L_{-2} +\tfrac{21}{5} (\s_2- \sigma_3^2\psi_0^2 -\s_3 \psi_1)J_{-2} {-3\s_3 (J_{-1}J_{-1}-J_0 J_{-2})}\nonumber \\
[J_{-1},W^4_{-1}]&=3 W^3_{-2}-2 \s_3\psi_0 L_{-2} +\tfrac{6}{5} (\s_2- \sigma_3^2\psi_0^2-\s_3 \psi_1)J_{-2}{-\s_3 (J_{-1}J_{-1}-J_0 J_{-2}) }\ . \nonumber
\end{align}
If $J_m$ and $W^4_n$ are the modes of local quasi-primary operators, it follows from the 
general analysis of \cite{Blumenhagen:1990jv} that their commutator has to take the form
\begin{align}
[J_{m,} W^4_n]  = & -3 m W^3_{m+n} -  \sigma_3\, \psi_0\, \tfrac{m}{2} \, (3m+n) L_{m+n}  + 
 \tfrac{m}{2} \, (3m+n) \Lambda^{(2)}_{m+n} \nonumber \\
& - (\s_2-\psi_0^2 \sigma_3^2-\s_3 \psi_1) \, \tfrac{m}{10} (5m^2 +5mn +n^2+1) J_{m+n}  \ , 
\end{align}
where $\Lambda^{(2)} = :JJ:$ is the normal ordered product of the spin-one current with itself. 
(In principle, also a normal ordered field at $s=3$ could have appeared, but this does not seem to be the case.) 
In any case, the bilinear contribution of the $J$-modes is not of the correct form --- in particular, one would 
have expected an infinite
sum of bilinear $J$ modes, and the coefficient in front of it does not have the correct $(m,n)$ dependence --- 
and it therefore follows that the $W^4$ modes cannot be the modes of a local field. In order to correct for this 
we define, for $|n| \leq 3$, 
\be\label{tildeW4def}
 \tilde{W}^4_n =  \frac{1}{2} \sum_{l} |3l-n| \, :L_{n-l}\,J_l: \, +\, \frac{1}{5} (3-2n^2) \Theta(2-|n|) \,J_0 L_{n} \ . 
\ee
For $|n| \leq 3$, these modes then satisfy the quasi-primary condition, 
\be
[L_m,\tilde{W}^4_n] =(3m-n) \tilde{W}^4_{m+n}\ ,\qquad m=0,\pm 1\ .	\label{tildeW4quasi}
\ee
The modified 
$s=4$ generators 
\be\label{V4def}
V^4_m=W^4_m - \s_3 \tilde{W}^4_m \ , 
\ee
then satisfy for $|n|\leq 3$ and all $m\in\mathbb{Z}$, 
\be\label{JV4}
 [J_m,V^4_n] 
 =-3m V^3_{m+n}- \tfrac{\s_2-\psi_0^2 \sigma_3^2}{10} \, m  (5 m^2+5 m n+n^2+1 )J_{m+n}\ .
\ee
Their commutators with the Virasoro modes can be similarly determined to be
\be
[L_m,V^4_n] =(3m-n) V^4_{m+n}-\tfrac{m(m^2-1)}{10}(7\s_2+3\s_3^2\psi_0^2)L_{m+n}\ ,
\label{LV4}
\ee
where we have again assumed $|n|\leq 3$. Both of these commutators are then of the local form predicted 
by \cite{Blumenhagen:1990jv}.

It should also be possible to extend the local modes $V^4_n$ beyond the wedge, i.e.\ for $|n|\geq 4$. However,
we have not found a closed formula for these modes. In any case, for the determination of the relevant 
structure constant, it suffices to consider the modes inside the wedge, see the discussion in Section~\ref{sec:sc}.

\bibliographystyle{JHEP}

\end{document}